\documentclass[12pt]{article}
\pdfoutput=1
\usepackage{jheppub}

\usepackage{graphicx}
\usepackage{textcomp} % for \textinterrobang
\usepackage{mathtools}
\usepackage{bm} % for \bm bold math command
\usepackage{mathrsfs} % for \mathscr bold math command

\usepackage{epstopdf}

\usepackage{latexsym}

\usepackage[mathcal]{euscript}
\usepackage{slashed}	

\usepackage{epic}
\usepackage{indentfirst}

\numberwithin{equation}{section}

\hypersetup{
	unicode,	% use with \texorpdfstring
	colorlinks,
	citecolor=[rgb]{0,.3,.5}, %cyan, % filecolor=black,% 
	linkcolor=[rgb]{0,.3,.5},
	urlcolor=[rgb]{0,.3,.5}, %cyan,% pdftex
	bookmarksopen=true,
	bookmarksopenlevel=\maxdimen,
}

\def\un#1{\underline #1}
\def\ul#1{\underline #1}
\def\vev#1{{\langle #1\rangle}}

\def\on#1#2{{\buildrel{\mkern2.5mu#1\mkern-2.5mu}\over{#2}}}

\def\dt#1{\smash{\on{\hbox{\bf .}}{#1}}\vphantom{#1}}     % Rev 2: make it same height as undotted symbol

\mathcode`\*="702A                  % * now always complex conjugate
\def\half{{\textstyle{1\over{\raise.1ex\hbox{$\scriptstyle{2}$}}}}}

\title{
Five-dimensional Supergravity in $N=1/2$ Superspace
} %end title

\author[a]{Katrin Becker,}
\author[a]{Melanie Becker,}
\author[a]{Daniel Butter,}
\author[a]{William D. Linch III,}
\author[b]{and Stephen Randall}
\affiliation[a]{
George P. and Cynthia Woods
Mitchell Institute for 
Fundamental Physics and Astronomy, \\
Texas A\&{}M University.\\
College Station, TX 77843, USA
} %end affiliation
\affiliation[b]{
Department of Physics, \\
University of California, \\
Berkeley, CA 90210, USA
} %end affiliation

\emailAdd{kbecker@physics.tamu.edu}
\emailAdd{mbecker@physics.tamu.edu}
\emailAdd{dbutter@tamu.edu}
\emailAdd{wdlinch3@gmail.com}
\emailAdd{stephenlrandall@gmail.com}

\preprint{MI-TH-191
}

\abstract{
We construct 5D, $N=1$ supergravity in a 4D, $N=1$ superspace with an extra bosonic coordinate. 
This represents four of the supersymmetries and the associated Poincar\'e symmetries manifestly. 
The remaining four supersymmetries and the rest of the Poincar\'e symmetries are represented linearly but not manifestly. 
In the linearized approximation, the action reduces to the known superspace result. 
As an application of the formalism, we discuss the construction of the 5D gravitational Chern-Simons invariant $\int A\wedge R\wedge R$ in this superspace.}

\begin{document}
\maketitle

%%%%%%%%%%%%%%%%%%%%%%%%%%%%%%%%%%%%%%%%%%%%%%%%%%%%%%%%%%%%%%%%
%%%%%%%%%%%%%%%%%%%%%%%%%%%%%%%%%%%%%%%%%%%%%%%%%%%%%%%%%%%%%%%%
\section{Introduction}

Five-dimensional, $N=1$ supergravity is a theory described in components by a frame $e_{\bm m}{}^{\bm a}(x)$, 
% a gravitino $\psi_{\bm m}^{\bm \alpha}(x)$, 
a gravitino $\psi_{\bm m}(x)$, and a graviphoton $A_{\bm m}(x)$ needed to match the bosonic degrees of freedom to the four fermionic degrees of freedom on-shell
\cite{Cremmer:1980gs,
Chamseddine:1980sp}. 
They are permuted by local supersymmetry,
but these transformations do not close onto translations and gauge transformations unless the equations of motion are imposed ({\it i.e.}\ they close only on-shell).
We can attempt to remedy this by introducing additional fields (the auxiliary component fields) and modifying the supersymmetry transformations in such a way that the algebra closes on all these fields off-shell. 
In this particular case, this off-shell problem is solved in a finite number of steps \cite{Breitenlohner:1978rq, Howe:1981ev}.\footnote{
The first off-shell construction of 5D supergravity was carried out by Zucker \cite{Zucker:1999ej, Zucker:1999fn}. Tensor calculus techniques for general 5D supergravity-matter actions were developed in \cite{Kugo:2000af, Fujita:2001kv, Bergshoeff:2002qk, Bergshoeff:2004kh}. The off-shell superspace appropriate for 5D was constructed in \cite{Kuzenko:2007cj, Kuzenko:2007hu}. A comprehensive discussion can be found in \cite{Butter:2014xxa}.}
But this is not so in general and fails even for this case when the theory is coupled to hypermultiplets.\footnote{See \cite{Siegel:1981dx} for the argument that generic hypermultiplets require an infinite number of auxiliary fields.}

When all of the supersymmetry is kept manifest, this state of affairs may be understood from the existence of off-shell superspaces with eight supercharges, of either the
harmonic \cite{Galperin:1984av,Kuzenko:2005sz}
or projective type \cite{Lindstrom:1989ne,Kuzenko:2007hu}.
% In these formalisms, superfields are defined on $SU(2)_R/U(1)$ resulting in an infinite expansion in the 2-sphere harmonics with ordinary superfields as coefficients.
Both approaches employ an auxiliary $SU(2) / U(1)$ space.
In the harmonic approach, superfields are globally defined on this space with an infinite expansion in the 2-sphere harmonics with ordinary superfields as coefficients. In the latter, superfields are instead holomorphic functions with infinite Laurent expansions in the natural inhomogeneous coordinate $\zeta$
of $\mathbb CP^1$.
% All of these are needed to close the supersymmetry off-shell, but only a small finite number of them survive when the equations of motion are imposed.
In both cases, all of these fields are necessary to close the supersymmetry off-shell, but only a small finite number of them survive when the equations of motion are imposed.  
% To reduce to a more familiar set of variables, one could, in principle, fix all but a finite amount of the harmonic symmetry, expand the superfields in half the $\theta$ variables, and integrate over harmonics. 
% The result would be an equivalent description in terms of a finite number of superfields depending only on the four remaining $\theta$s but on all five bosonic coordinates. 
% These fields behave as 4D, $N=1$ superfields in almost every way. 
To reduce to a more familiar set of variables, one could, in principle, eliminate all but a finite number of fields (by solving auxiliary field equations and imposing gauge symmetries), expand the superfields in half the $\theta$ variables, and integrate over the auxiliary manifold. 
%This is a natural procedure in projective superspace, where the parametrization selects out the polar axis corresponding to an $N=1$ subspace.
The result would be an equivalent description in terms of a finite number of superfields depending only on the four remaining $\theta$'s but on all five bosonic coordinates. 
These fields behave as 4D, $N=1$ superfields in almost every way.

% The rub lies in the ``in principle'' hedge, as this procedure appears to be exceedingly difficult to carry out explicitly.\footnote{To our knowledge, there is no algorithm to reduce general models in 4D, $N=2$ superspaces to 4D, $N=1$ explicitly. A sense of what is involved may be gotten from \cite{Butter:2010sc}. There 4D, $N=2$ supergravity in projective superspace is linearized around flat space, then partially gauge-fixed, integrated over harmonics, and put partially on-shell to relate it to the 4D, $N=1$ old- and new-minimal supergravity theories.}

This procedure is difficult to carry out explicitly in supergravity.\footnote{
% To our knowledge, there is no algorithm to reduce general models in 4D, $N=2$ superspaces to 4D, $N=1$ explicitly. 
A sense of what is involved may be gotten from \cite{Butter:2010sc}. There 4D, $N=2$ supergravity in projective superspace is linearized around flat space, then partially gauge-fixed, integrated over harmonics, and put partially on-shell to relate it to the 4D, $N=1$ old- and new-minimal supergravity theories. To incorporate supergravity fully non-linearly, one would actually need to repeat the procedure of \cite{Banerjee:2011ts} in reducing 5D, $N=1$ to 4D, $N=2$ (but keeping dependence
on the fifth coordinate and in superspace) and then recast 4D, $N=2$ derivatives into 4D, $N=1$ language
(see e.g. \cite{Labastida:1984qa, Labastida:1986md}).}
Worse still, in dimensions higher than six and/or for more than eight Poincar\'e supercharges, there are no appropriate off-shell superspaces over which we could even contemplate carrying out such a procedure \cite{Berkovits:1993hx}.\footnote{``Appropriate'' here means, roughly, that it is possible to separate the constraints defining linear, off-shell, irreducible representations from the equations of motion that put them on-shell.}
Instead, one could start from a set of superfields and transformation rules that {\em would have} resulted from the purported procedure and attempt to construct an invariant action directly.
Specifically, we first embed the component fields into a suitable 1-parameter family of 4D, $N=1$ superfields with the parameter having the interpretation of a coordinate for the  5$^\textrm{th}$ dimension.
The embedding will be suitable if the component spectrum and all gauge transformations are reproduced. 
One then attempts to construct an action from the superfield ingredients that is invariant under the superfield gauge transformations.

Such a ``superfield Noether procedure'' is guaranteed to be possible when the off-shell superspace exists and, indeed, it was used to construct five- and six-dimensional matter-coupled supergravity in various approximations in \cite{Linch:2002wg, Buchbinder:2003qu, Gates:2003qi,
Paccetti:2004ri, Abe:2005ac, Sakamura:2012bj, Sakamura:2013cqd,
Abe:2015bqa, Abe:2015yya, Abe:2017pvw}.
In the cases in which an off-shell superspace does not exist, it is not obvious that the procedure will work. Nevertheless, it was explicitly shown that it does work for 10D super-Yang-Mills in \cite{Marcus:1983wb}.
More recently, the approach was extended to the far more subtle case of eleven-dimensional supergravity in \cite{Becker:2016xgv, %ATH
Becker:2016rku,	%NATH
Becker:2016edk,	%G2Potential
Becker:2017njd, 	%All CS
Becker:2017zwe,	%Linearized
Becker:2018phr}.	%Current

In this paper, we revisit the five-dimensional problem with the benefit of eleven-dimensional hindsight. 
Our motivations for doing this are both five-dimensional and eleven-dimensional. 
On the one hand, the existing five-dimensional results are only understood in various approximate forms whereas the analytic structure of the eleven-dimensional action is now known to much higher precision. 
On the other hand, the five-dimensional version is far simpler than its eleven-dimensional counterpart while retaining many of the non-trivial elements of the latter. 
Specifically, both are odd-dimensional and have an abelian $p$-form in their on-shell spectra ($p=1$ and $3$, respectively) with a Chern-Simons-like self-interaction. 
(In fact, it was noted already in \cite{Linch:2002wg} that the five-dimensional theory also has a component 3-form hiding in its scale compensator so that, in the end, the two theories employ almost identical 4D, $N=1$ superfield representations.)
Finally, both theories get higher-derivative corrections involving the Chern-Simons form and powers of the curvature 2-form. 
In 5D, the supersymmetric completion is known \cite{Hanaki:2006pj, Bergshoeff:2011xn}, whereas in 11D only parts of it have been constructed. 

In the next section, we will give the embedding of the component fields of five-dimensional supergravity into a 1-parameter family of 4D, $N=1$ superfields. 
The parameter is the 5$^\textrm{th}$ bosonic coordinate, so we are describing 5D, $N=1/2$ superspace as a fibration by 4D, $N=1$ superspaces over the fifth dimension.
Starting from the superfields  on this space and their gauge transformations, we construct their field strengths and the Bianchi identities they obey. 
In section \ref{S:Action}, we define a perturbation theory in the gravitino and give the action to the first non-trivial order in this expansion. 
The result to lowest order is the sum of a D-term integral representing the volume of our superspace and a mixed F- and D-term Chern-Simons action for the graviphoton. 
In section \ref{S:1DSpinorGeometry}, we explain the structure of our result by deriving it from ordinary on-shell, five-dimensional superspace.
In section \ref{S:LinearizedAction}, we linearize our action to recover the previously known superspace action for linearized supergravity \cite{Linch:2002wg}.
%As an application of our formalism, we address i
In section \ref{S:R2} we apply the formalism to the gravitational Chern-Simons term $\sim \int  A\wedge R \wedge R$ where $A$ is the graviphoton and $R$ is the curvature 2-form. 
In particular, we construct the minimal $N=1$ sector containing the 4D $R \wedge R$, which is non-trivial due to the competing requirements of chirality and gauge-invariance. 
%\added
{(The construction is in terms of linearized field strengths but is, at least in part, expected to hold beyond this order.)}
We conclude in section \ref{S:End} with a summary of our findings and a discussion of implications for extensions and future work. 
In the main thread of our presentation, we have opted to suppress distracting calculations for the sake of clarity, relegating them instead to appendix \ref{S:Derivations}.

\paragraph{Note added in this revised version:} 
In this revised version, we have added the clarifying sections \ref{S:Formulations} and \ref{S:Discussion}, expanded the discussion in section \ref{S:1DSpinorGeometry}, and added a new section \ref{S:NewMinimal}.
In \S{}\ref{S:Formulations} we review the existing formulations of off-shell 5D, $N=1$ Weyl and Poincar\'e supergravity theories and explain the connection to the 5D, $N=1/2$ superspace formulation. This helps to elucidate the structure of the non-manifest and on-shell part of the supersymmetry. 
%This gives some insight into higher-dimensional extensions as well as theories with more supersymmetry for which no off-shell formulation is known. 
With this understood, we can see that there is an alternative linearization of our results, which we present in the new
section \ref{S:NewMinimal}. 
In the new section \S{}\ref{S:Discussion}, we compare our gravitational Chern-Simons action to results available in the literature. 
These and other clarifying remarks were added in response to thoughtful questions from our JHEP referees.

%%%%%%%%%%%%%%%%%%%%%%%%%%%%%%%%%%%%%%%%%%%%%%%%%%%%%%%%%%%%%%%%
%%%%%%%%%%%%%%%%%%%%%%%%%%%%%%%%%%%%%%%%%%%%%%%%%%%%%%%%%%%%%%%%
\section{Fields and Symmetries}
\label{S:FieldsSymmetries}

\paragraph{Component fields}
In 4D, $N=1$ supergravity, the component content is that of a frame $e_m{}^a$ 
% and its superpartner gravitino $\psi_m^\alpha$.
and its superpartner gravitino $\psi_m$.
This simple matching of bose and fermi degrees of freedom is not a generic trait; 
counting the degrees of freedom of a five-dimensional frame field $e_{\bm m}{}^{\bm a}$ 
% and gravitino $\psi_{\bm m}^{\bm \alpha}$,
and gravitino $\psi_{\bm m}$, 
we conclude that the 5D, $N=1$ supergravity multiplet is missing some bosonic fields.
On shell, the bose and fermi degrees of freedom can be matched by a five-dimensional ``graviphoton'' gauge vector $A_{\bm m}$: The 5D, $N=1$ supergravity component multiplet is 
% the set $\{ e_{\bm m}{}^{\bm a} , \psi_{\bm m}^{\bm \alpha} , A_{\bm m}\}$ 
the set $\{ e_{\bm m}{}^{\bm a} , \psi_{\bm m}, A_{\bm m}\}$ 
transforming into each other under linearized supersymmetry transformations.

%%%%%%%%%%%%%%%%%%%%%%%%%%%%%%%%%%%%%%%%%%%%%%%%%%%%%%%%%%%%%%%%
\subsection{Superfields}
To embed these fields into representations of the 4D, $N=1$ super-Poincar\'e algebra, we split their polarizations  along the 4D directions $x^m$ and the extra dimension $y=x^5$.
Using four-dimensional language, this results in a graviton $e_m{}^a$, a KK-photon $\mathcal A_m \sim e_m{}^5$, a scalar $\varphi \sim e_y{}^5$, a graviphoton $A_m$, a pseudoscalar $A_y$, two gravitini, and two gaugini, but with all these fields depending on all five bosonic coordinates $x^m$ and $y$. 
That these can be properly accommodated in 4D, $N=1$ superfields was originally demonstrated in \cite{Linch:2002wg, Gates:2003qi} in a linearized approximation. 
Here, we will use a larger set of superfields from the outset as this is more geometric and thus convenient when constructing the non-linear theory:
\begin{description}
\item[Conformal supergravity] 
\label{L:huh}
We embed the 4D polarizations of the frame $e_m{}^a$ and one gravitino into the conformal 4D, $N=1$ supergravity prepotential $U^a(x, \theta, \bar \theta, y)$ \cite{Gates:1983nr, Buchbinder:1998qv}. This field has a large gauge freedom, the linearized part of which is\footnote{Here and henceforth, we use the convention that a vector index, $v^a$ say, that is contracted on a Pauli matrix is denoted by an underline: $v^{\un a} := v^{\alpha \dt \alpha}:=(\sigma_a)^{\alpha \dt \alpha} v^a $. (This is essentially the Feynman slash notation but on the index instead of the vector, which proves to be more convenient in superspace calculations.) An implication of this is that contracted underlined indices give traces of Pauli matrices so that, for example, $v^{\un a} \eta_{\un a} = -2 v^a \eta_a$. 
}
\begin{align}
\label{E:GravitonLinearized}
\delta_0 U^{\un a} = \bar D^{\dt \alpha} L^\alpha - D^\alpha \bar L^{\dt \alpha} 
~.
\end{align}
The arbitrary spinor $L^\alpha(x, \theta, \bar \theta, y)$ allows a Wess-Zumino gauge in which the only non-zero components are the frame, the gravitino, and a real pseudo-vector auxiliary field. The remaining gauge parameters are those of linearized diffeomorphisms, local supersymmetry, local Lorentz, local S-supersymmetry, and local scale. The latter two can be used to remove the spin-1/2 and spin-0 parts of the supergraviton, so this multiplet describes conformal supergravity instead of ordinary Poincar\'e supergravity. We will make up for this presently by introducing a scale compensator superfield.

\item[Conformal gravitino] The other gravitino is embedded into the conformal gravitino superfield $\Psi^\alpha$. Again this representation has a large gauge freedom transforming in the linearized approximation as 
\begin{align}
\label{E:GravitinoLinearized1}
\delta_0 \Psi^\alpha = \Xi^\alpha + D^\alpha \Omega
\end{align}
for a chiral spinor $\Xi^\alpha$ and complex scalar superfield $\Omega$ \cite{Gates:1983nr}. 
These can be used to go to a Wess-Zumino gauge where
\begin{align}
%\label{E:GravitinoLinearized1a}
\Psi^\alpha \sim \cdots 
	+ i\theta \sigma^m \bar \theta \psi_m{}^\alpha 
	+ \theta^2 (\sigma_m\bar \theta )^\alpha v^m 
	+ \bar \theta^2 (\theta \sigma^{mn})^\alpha t_{mn}
	+ \theta^2\bar \theta^2 \rho^\alpha
~.	
\end{align}
In addition to linearized supersymmetry, the conformal gravitino $\psi_m{}^\alpha$ possesses a shift symmetry in its spin-1/2 part corresponding to local S-supersymmetry.
The additional fields are auxiliary and consist of a complex vector $v^m$, a complex self-dual 2-form $t_{mn}$, and a spinor $\rho^\alpha$. Note that there is no physical boson remaining in this set.\footnote{This is not in direct conflict with supersymmetry, because there is no single-derivative Lagrangian for this representation unless it is coupled to other fields. (There is a higher-derivative Lagrangian, but then an otherwise-auxiliary vector becomes dynamical.)}

\item[Kaluza-Klein gauge field] The mixed component of the metric is described by a non-abelian connection for diffeomorphisms in the extra $y$-direction. This is implemented in superspace by covariantizing the flat superspace derivatives $D\to \mathcal D$. The non-abelian field strength appears in the derivative algebra in the usual place \cite{Gates:1983nr, Buchbinder:1998qv, Wess:1992cp}
\begin{align}
\label{E:NonAbConnex}
[\bar {\mathcal D}_{\dt \alpha} , \mathcal D_{b} ] =  (\sigma_b)_{\alpha \dt \alpha} \mathcal L_{\mathcal W^\alpha}
\end{align}
where for any vector field $v^y\partial_y$, $\mathcal L_v $ denotes the Lie derivative along $v$. 
(The Lie derivative appears here because this field gauges the diffeomorphisms in the $5^\textrm{th}$ dimension.)
As usual, the derivative constraints can be solved in term of a non-abelian prepotential $\mathcal V^y$.\footnote{The
sign convention for $\mathcal V^y$ and $\mathcal W_\alpha{}^y$ differs from our recent papers in 11D.}

\item[Graviphoton hierarchy] The graviphoton has a part embedded into an abelian vector prepotential $V$ with the usual gauge transformation $\delta V = \tfrac1{2i}(\Lambda - \bar \Lambda)$ allowing the standard Wess-Zumino gauge. The other part $A_y$ is carried by a chiral pseudo-scalar $\Phi_y$ transforming into the same chiral gauge parameter $\delta \Phi_y = \partial_y \Lambda$. 
This is a short abelian tensor hierarchy in which only a vector multiplet and a scalar multiplet are linked. 
Below, we will ``gauge it'' by defining $\Phi$ (and, therefore, $\Lambda$) to be covariantly chiral under the nonabelian connection $\mathcal D$.
%\end{description}

\item[Gauge 3-form compensator]
At this point it appears we have embedded all of the component fields of 5D, $N=1$ supergravity.
We must recall, however, that the gauge transformation \eqref{E:GravitonLinearized} of the graviton superfield removes the spin-1/2 and spin-0 component fields. To compensate for this gauging of (super)scale transformations, one introduces a superfield representation that contains a scalar. 
The standard choice is to introduce a chiral scalar superfield \cite{Siegel:1978mj}.
More appropriate to our case, however, is a {\em constrained} chiral scalar $G = -\tfrac14 \bar D^2 X$ with $\bar X=X$ a real scalar prepotential transforming under linearized superconformal transformations as $\delta_{sc} X = D^\alpha L_\alpha  + \bar D_{\dt \alpha} \bar L^{\dt \alpha}$ \cite{Linch:2002wg, Becker:2016edk}.
%\begin{align}
%\delta_{sc} X = D^\alpha L_\alpha  + \bar D_{\dt \alpha} \bar L^{\dt \alpha}~.
%\end{align}
This rule is needed so that $G$ transforms as the conformal compensator $\delta_{sc} G = -\tfrac14 \bar D^2 D^\alpha L_\alpha$ under linearized transformations.
Note, however, that the chiral part of $L^\alpha$ leaves $G$ invariant, so there is a gauge-for-gauge symmetry under which $L^\alpha \to L^\alpha+ \tfrac 1{2i} \Upsilon^\alpha$, where $\Upsilon^\alpha$ is a chiral spinor parameter. 
Because of this symmetry, the complex F-term auxiliary field survives in Wess-Zumino gauge and furthermore, since $X$ is real, 
%instead of a complex F-term auxiliary field, 
the imaginary part of $G$'s auxiliary field is the dual of a 4-form field strength \cite{Gates:1983nr, Gates:1980ay,Gates:1980az}. 
Equivalently, the $\theta \sigma^m\bar \theta$ component of the $X$ prepotential is the Hodge dual of a gauge 3-form $C_{mnp}$.
(That this is a gauge 3-form can be derived from the gauge transformation $\delta X= \tfrac1{2i}(D^\alpha \Upsilon_\alpha - \bar D_{\dt \alpha} \bar \Upsilon{}^{\dt \alpha})$.)

In the eleven-dimensional theory, this 3-form is the M-theory 3-form with all legs in the four-dimensional directions \cite{Becker:2016edk}.
In the five-dimensional case, the geometrical origin of this form is less apparent, but we will see that there is a complete super-3-form of compensating fields needed for consistency. 
In anticipation of this, we introduce the gauge chiral spinor superfield $\Sigma^\alpha_y$ transforming under the abelian 1-form symmetry as $\delta \Sigma^\alpha_y = -\tfrac 14 \bar D^2 D^\alpha u_y + \partial_y \Upsilon^\alpha$.
Together, $X$ and $\Sigma^\alpha_y$ form a second short tensor hierarchy describing a five-dimensional gauge 3-form $C_{\bm{mnp}}$ in a $4+1$ split $C_{mnp}$ and $C_{mn\,y}$ \cite{Becker:2016xgv}. 
\end{description}

%%%%%%%%%%%%%%%%%%%%%%%%%%%%%%%%%%%%%%%%%%%%%%%%%%%%%%%%%%%%%%%%%%%%%%%%%%%%%%%%%%%%%%%%%%%%%%%%
\subsection{Non-abelian tensor hierarchy}
Previously, we gauged the 1-form hierarchy by replacing $D\to \mathcal D$. 
In fact, this couples the KK vector field correctly to all fields, since it builds the non-abelian correction directly into the superspace geometry \eqref{E:NonAbConnex}. 
This gives rise to corrections to the Bianchi identities for closed $p$-forms, and, therefore, to the field strengths and gauge transformations.
Explicitly, the Bianchi identities for a closed 2-form are
\begin{subequations}
\label{E:NATHBI}
	\begin{align}
	\label{E:NATHBIFW}
	-\frac14 \bar {\mathcal D}^2 {\mathcal D} F_y + \partial_y W  &= 0
	\\
	{\mathcal D}^\alpha W_\alpha - \bar {\mathcal D}_{\dt \alpha} \bar W^{\dt \alpha}  &= -2i \omega(\mathcal W^y , F_y) 
	\end{align}
with $\omega(\mathcal W^y , F_y) := \mathcal W^{\alpha y} \mathcal D_\alpha F_y +\tfrac12 \mathcal D^\alpha \mathcal W^y_\alpha F_y +\mathrm {h.c.}$, and
	\begin{align}
	\label{E:NABI}
	{\mathcal D}^\alpha \mathcal W^y_\alpha - \bar {\mathcal D}_{\dt \alpha} \bar {\mathcal W}^{\dt \alpha y} =0 
	~.
	\end{align}
The 4-form satisfies 
	\begin{align}
	\label{E:NATHBIH}
	-\frac14 \bar {\mathcal D}^2  H_y + \partial_y G &=0
	\\
	\label{E:NATHBIG}
	\bar {\mathcal D}_{\dt \alpha} G &=0
	\end{align}
\end{subequations}
with the {\it proviso} that $G$ has a real prepotential ({\it i.e.}\ one of its auxiliary fields is the dual of a 4-form field strength).
This non-abelian tensor hierarchy of constraints is solved by the field strengths
\begin{subequations}
\label{E:NATHFS}
\begin{align}
\label{E:NATHFS1}
F_y &:= \frac1{2i} (\Phi_y - \bar \Phi_y) - \partial_y V
\\
W^\alpha &:= -\frac14 \bar {\mathcal D}^2{\mathcal D}^\alpha V - {\mathcal W}^{\alpha y} \Phi_y
\\
\label{E:NATHFS3}
H_y &:= \frac1{2i}\left(\mathcal D^\alpha \Sigma_{\alpha y} - \bar {\mathcal D}_{\dt \alpha} \bar \Sigma_y^{\dt \alpha} \right) - \partial_y X
\\
G &:= -\frac14 \bar {\mathcal D}^2 X - \mathcal W^{\alpha y} \Sigma_{\alpha y}
\end{align}
\end{subequations}
in terms of unconstrained prepotentials. 
These expressions are the standard ones of the abelian hierarchy with the minimal coupling prescription $D\to \mathcal D$ and corrections by the non-abelian field strength.

The prepotentials, in turn, suffer pre-gauge transformations 
\begin{subequations}
\label{E:NATHgauge0}
\begin{align}
\label{E:NATHgauge0Phi}
\delta_0 \Phi_y &= {\mathcal L}_{\lambda} \Phi_y+  \partial_y \Lambda
\\
\label{E:NATHgauge0V}
\delta_0 V &= {\mathcal L}_{\lambda} V + \frac1{2i} (\Lambda- \bar \Lambda) 
\\
\delta_0 \Sigma^\alpha_y &= {\mathcal L}_{\lambda} \Sigma^\alpha_y - \frac1{4} \bar {\mathcal D}^2 {\mathcal D}^{\alpha} u_y + \partial_y \Upsilon^\alpha
\\
\label{E:delta0X}
\delta_0 X &= {\mathcal L}_{\lambda} X 
	+ \frac1{2i} \left[ 
		{\mathcal D}^\alpha \Upsilon_\alpha
		- \bar {\mathcal D}_{\dt \alpha} \bar \Upsilon^{\dt \alpha}
		\right]
	- \omega(\mathcal W^y, u_y)
~.
\end{align}
\end{subequations}
The first term is the non-abelian part of the gauge transformation (corresponding to diffeomorphisms of the circle) which acts on matter fields by the Lie derivative. 
The remaining terms are minimal covariantizations $D\to \mathcal D$ of the abelian $p$-form transformations and, in the case of $X$, a correction term needed to counter the appearance of a non-abelian field strength in the commutator of four $\mathcal D$'s.
The field strengths \eqref{E:NATHFS} are invariant under the abelian part of the transformations and are covariant under the non-abelian part \cite{Becker:2016rku}.

In addition, all of the prepotentials transform under $L_\alpha$ transformations (see {\it e.g.}\
section 5 of \cite{Becker:2018phr}). The relevant ones for our discussion will be the $L_\alpha$
transformations of $\Sigma_y^\alpha$ and $X$, which take the form
\begin{align}\label{E:LalphaTrafo}
\Delta \Sigma_{y}^{\alpha} = \frac{i}{2} \bar{\mathcal D}^2 (L^\alpha H_y)
~~~\textrm{and}~~~
\Delta X = \mathcal D^\alpha (L_\alpha G) - i L^\alpha \mathcal W_\alpha{}^y H_y + \text{h.c.}
\end{align}
for covariantized transformations $\Delta$ defined in \cite{Becker:2018phr}.
In the linearized background where $\vev G=1$, the $L^\alpha$ transformation of $X$ does not
vanish, which identifies it as the prepotential for the conformal compensator.

%%%%%%%%%%%%%%%%%%%%%%%%%%%%%%%%%%%%%%%%%%%%%%%%%%%%%%%%%%%%%%%%%%%%%%%%%%%%%%%%%%%%%%%%%%%%%%%%
\subsection{Gravitational couplings}
\label{E:GrinoPunt}

In order to include the full non-linear couplings to gravity, one should introduce a
gravitational covariant $\mathcal D\to \mathscr D$, whose connections include the supervielbein,
Lorentz connection, etc. in addition to the non-abelian gauge field.  As discussed in
\cite{Becker:2018phr}, for this to be consistent with $y$-dependence of the 4D supergraviton, we will be required to incorporate also the gravitino superfield corrections to the supergeometry: The graviton transformation \eqref{E:GravitonLinearized} must be allowed to depend on $y$, so we expect terms $\sim{\partial_y} L^\alpha$ to appear that we need to be able to cancel. This can be done provided we modify the gravitino superfield transformation \eqref{E:GravitinoLinearized1} by a term $\sim{\partial_y} L^\alpha$ \cite{Linch:2002wg, Becker:2017zwe}.\footnote{A more covariant version of this statement can be made by studying the consistency of the Bianchi identities of the commutator $[\mathscr D_\alpha, \mathscr D_y]$.}
This gravitino superfield, in turn, contributes its torsions and curvatures to the covariant derivative algebra \cite{Gates:1984mt, Gates:1983nr}, which would first need to be constructed. 
This approach has been carried out for general superspaces arising in Kaluza-Klein splittings of the type we are considering and will be reported separately. 

Instead, we here take the far simpler approach of defining a gravitino expansion and working order-by-order in $\Psi^\alpha$. 
In this approach, we can still work in a non-linear 4D, $N=1$ conformal supergravity background provided that background is $y$-independent. 
Even in this setting the dependence on the remaining fields is non-linear. 
Explicitly, we treat only $U^a$ and $\Psi_y^\alpha$ perturbatively in their $y$-dependence.
This can be done around any $y$-independent 4D, $N=1$ conformal supergravity background, such as warped compactifications over 4D solutions. (For example, we can describe $AdS_5$ in the Poincar\'e patch as a warped Minkowski compactification.)
For this reason, we will write the 4D, $N=1$ conformal supergravity measures explicitly in the rest of this section and in the next.
Covariant derivatives are understood to be background-covariant derivatives. Formally this amounts to replacing $\bar {\mathcal D}^2 \to \bar {\mathcal D}^2 - 8 R$ wherever they appeared previously. (No other torsions can appear at this dimension.)
We emphasize that this restriction applies only to the invariants of 4D, $N=1$ conformal supergravity part and that this restriction can be removed by using the more complicated supergeometry alluded to above.

\paragraph{Local superconformal symmetry}
Separating the components of 5D, $N=1$ gravity in a $4+1$ split and embedding the 4D polarizations of the frame in a 4D, $N=1$ {\em conformal} supergraviton \eqref{E:GravitonLinearized} has resulted in a description with a local 4D superconformal symmetry. 
This symmetry is broken spontaneously to Poincar\'e by the compensators just as 4D, $N=1$ Poincar\'e supergravity is usually described by conformal superframes coupled to scale compensators \cite{Siegel:1978mj,Gates:1983nr}.
Similarly to that case, the matter fields can be assigned scaling weights $\Delta$ and $U(1)$ charges $w$ in addition to their engineering dimension $d$ and degree $q$ as differential forms on $Y$. 
These are collected in table \ref{T:Superscale}.

\begin{table}[t]
\begin{align*}
%\hspace{-10mm}
{\renewcommand{\arraystretch}{1.3} %adds some padding
\begin{array}{|c|rrrrrrrr|rrrrrrc|}
\hline
	& {\mathcal D}_\alpha &  U^a & \Psi^\alpha & G(X) & H(\Sigma) & W_\alpha(V) &  F(\Phi) & \mathcal W(\mathcal V) 
	& L_\alpha & \Xi^\alpha & \Omega & \Upsilon_\alpha& u & \Lambda & {\partial_y}
\\
\hline
\Delta& \tfrac12	&  -1 & -\tfrac32 & 	3(2) 	& 2 (\tfrac32) & \tfrac32(0) & 0(0) & \tfrac32(0)
	& -\tfrac32& -\tfrac32   & 0 & \tfrac32 & 0 &  0& 0 
\\
w 		& -1		& 0 	& -1  &   2(0) 	&  0 (1) 	& 	1(0)	& 0(0)& 	1(0)
	&-1 		&-1 	& 0 & 1&	0& 0& 0
\\
d 			&\tfrac12	& -1	&- \tfrac12  &	0(-1)	&	0(-\tfrac12)& \tfrac12(-1) & 0(0)&\tfrac12(-1)
	&-\tfrac32&-\tfrac12& -1 & -\tfrac32& -2& -1& 1
\\
q & 0 & 0 & 1 & 0 & 1 & 0 &1 & -1
	& 0& 1 & -1 & 0 & 1  &0 & 1\\
\hline
\end{array}
}
\end{align*}
\caption{
The various $\mathbf Z$-gradings of the fields and gauge parameters:
scaling dimension $\Delta$, $U(1)_R$ weight $w$, mass dimension $d$, and 5-charge $q$
(with $q=p$ for $p$-forms, $q=-1$ for vectors, etc.)
}
\label{T:Superscale}
\end{table}

%%%%%%%%%%%%%%%%%%%%%%%%%%%%%%%%%%%%%%%%%%%%%%%%%%%%%%%%%%%%%%%%%%%%%%%%%%%%%%%%%%%%%%%%%%%%%%%%
\subsection{Gravitino perturbation theory}
Finally, we complete the description of the symmetries by formulating a perturbation theory in the gravitino. 
To do this, we introduce a gravitino grading under which $\Psi$ carries charge $+1$ and the remaining prepotentials have charge 0. 
Then the superconformal parameters under which the gravitino transforms \eqref{E:GravitinoLinearized1} must carry charge 1 as well. 
We split up the gauge transformations $\delta = \delta_{-1} + \delta_0 + \delta_1 + \dots$, and assign gravitino charges to all the fields and gauge parameters. 
The $\delta_0$ transformations are taken to be those defined on all the fields above.
The $\delta_{-1}$ transformation acts only on the gravitino as
\begin{gather}
\label{E:delta-1}
\delta_{-1} \Psi^\alpha_y :=  2i \partial_y L^\alpha
\end{gather}
This is needed to covariantize the $y$-dependence of the superconformal graviton transformations under $L^\alpha$, as described in the previous paragraph.
The $\delta_1$ transformations can act on the all the non-gravitino fields only by $\Xi$ and $\Omega$. 
The precise form of this action is not easy to derive from first principles but can be guessed and checked. 
In the process of bootstrapping, we are aided immensely by the huge amount of local symmetry represented in table \ref{T:Superscale}.
We find for $\delta_1$ that it acts on the tensor hierarchy fields as
\begin{align}
\label{E:NATHgauge1}
\delta_1  \Phi_y &= \Xi^\alpha_y W_\alpha 
~,~~
\delta_1 V = \check \Omega^y F_y
~,~~
\delta_1 \Sigma_y^\alpha = -G\Xi_y^\alpha 
~,~~
\delta_1 X = \check \Omega^y H_y
~,~~
\delta_1 \mathcal V^y = \hat \Omega^y
\end{align}
and by zero on all other fields. Here $\hat \Omega^y$ and $\check \Omega^y$ correspond to the
real and imaginary parts of the $\Omega$ parameter.

Note that these transformations are rather large, implying that many component fields can be removed by a choice of superscale gauge. 
Most apparently, the linearized non-abelian gauge field suffers a St\"uckelberg shift by an unconstrained real superfield, indicating that it can be gauged away entirely!
Similarly, since $G$ is the scale compensator superfield, the lowest bosonic component must be non-zero ({\it i.e.}\ we can gauge $G| \to 1$). 
It follows that the 2-form superfield $\Sigma_y^\alpha$ can also be gauged away completely. 
Finally, we will see in the next section that the lowest component of $F_y$ is the volume density on $Y$, so it must be non-vanishing and therefore invertible. 
Thus, $V$ is also a St\"uckelberg superfield shifting under the imaginary part of $\Omega$.
Where are all these fields going?
On its face, it strains credulity that we are able to remove them all, but this is, in fact, consistent with the description of linearized 5D, $N=1$ supergravity given in \cite{Linch:2002wg}:
As we will see in detail in section \ref{S:LinearizedAction}, the St\"uckelberg components are being eaten by the gravitino superfield in a supergravity Higgs-like mechanism. 
Fully ``massing up'' the gravitino superfield results in the superfield spectrum used in the aforementioned reference.

%%%%%%%%%%%%%%%%%%%%%%%%%%%%%%%%%%%%%%%%%%%%%%%%%%%%%%%%%%%%%%%%
%%%%%%%%%%%%%%%%%%%%%%%%%%%%%%%%%%%%%%%%%%%%%%%%%%%%%%%%%%%%%%%%
\section{Action}
\label{S:Action}
Having defined the fields and symmetries in low orders of the gravitino expansion, we are finally in a position to construct the action.
To this end, we expand the action $S = S_0+ S_1 +S_2 +\dots$ so that $\delta S=0$ splits up into a set of conditions with definite gravitino grading:
\begin{gather}
\label{E:GrinoExpansion}
\delta_0 S_0 + \delta_{-1} S_1 = 0
~,~~
\delta_1 S_0 + \delta_0 S_1 + \delta_{-1} S_2= 0
~,~~
\dots
\end{gather}
Note that, while this is an expansion in the gravitino superfield $\Psi$, each term is non-linear in all the remaining fields.
In the next two subsections, we will give the explicit exact results for $S_0$ and $S_1$ that satisfy these relations.

%%%%%%%%%%%%%%%%%%%%%%%%%%%%%%%%%%%%%%%%%%%%%%%%%%%%%%%%%%%%%%%%
%\subsection{Volume Term and Chern-Simons Action}
\subsection{Chern-Simons Action}
\label{S:CSA}

\begin{table}[t]
\begin{align*}
{\renewcommand{\arraystretch}{1.3} %adds some padding
\begin{array}{|c|rrrrrrrrc|}
\hline
	&\kappa^2& \int dy &\int d^4x & \int d^4 \theta E & \int d^2 \theta \mathcal E& \partial_a & {\partial_y} & W & K \\
\hline
\Delta &0& 0& -4& 2 & 1& 1 & 0 &  3& 2 \\
w &0& 0&0 & 0 &-2 & 0 & 0& 2&  0 \\
d  &-3& -1 &-4& 2& 1 & 1& 1& 1& 0\\
q  &0& -1 &0& 0& 0 & 0& 1& 1& 1\\
\hline
\end{array}
}
\end{align*}
\begin{caption}{
Weyl ($\Delta$) and $U(1)_R$ ($w$) weights of various measures and actions. $W$ and $K$ stand for the integrands of F- and D-term superspace integrals.
}
\label{T:ActionWeights}
\end{caption}
\end{table}

The lowest-order action 
\begin{align}
S_0 &= S_{vol} + S_{CS}
\end{align}
is a sum of terms separately invariant under $\delta_0$. 
The simplest of these is the Chern-Simons action $S_{CS} = \int d^4 x \int dy L_{CS\,y}$.
We write the Lagrangian in the slightly clumsy way (because this will be the form that we can generalize appropriately below):
\begin{align}
\label{E:CSLagrangian}
2 \kappa^2 L_{CS\, y} = i\int d^2 \theta \mathcal E\, \Phi_y \mathbb G 
	+ \int d^4 \theta  E\, V \mathbb H_y
	+ \textrm{h.c.} 
\end{align}
with
\begin{align}
\label{E:Composites1}
\mathbb G= \frac12 W^\alpha W_\alpha 
~~~\mathrm{and}~~~
\mathbb H_y = \omega(W, F_y)
	-i {\mathcal D}^\alpha F_y \mathcal W_\alpha^y F_y
	+i \bar {\mathcal D}_{\dt \alpha} F_y \bar {\mathcal W}^{\dt \alpha y} F_y
\end{align}
where, as before, $\omega(W, F) := W \mathcal D F + \frac 12 F \mathcal D W +\textrm{h.c.}$ is the Chern-Simons superfield.
This Lagrangian is a 1-form on $Y$ that can be understood as a superfield $[4,1]$-form on $\bm X \times Y$ \cite{Becker:2017njd}.
It is constructed in terms of a composite 4-form with $[4,0]$ part $\mathbb G$ and $[3,1]$ part $\mathbb H=\mathbb H_y\, dy$. 
As the notation is intended to suggest, they satisfy the same Bianchi identities (\ref{E:NATHBIH}, \ref{E:NATHBIG}) as $G$ and $H_y$. Specifically, the action is invariant under the $\Lambda$ transformation (\ref{E:NATHgauge0Phi} ,\ref{E:NATHgauge0V}) because $\bar {\mathcal D}^2 \mathbb H = 4 \partial_y \mathbb G$. 

\subsection{Volume Term and Chern-Simons Action}
The second invariant is the superspace volume
\begin{align}
\label{E:vol}
S_{vol} &:= - \frac3{\kappa^2} \int d^4x \int dy \int d^4\theta E \, (\bar G G)^{1/3} F_y \,{\mathcal F}
\end{align}
determined up to some completely weightless function ${\mathcal F}$ of the field strengths \eqref{E:NATHFS}.
The integrand of \eqref{E:vol} is again a 1-form on $Y$ that can be understood as the superspace volume density dressed up for conformal invariance:
The explicit factor of $F_y$ may be interpreted as an ein-bein superfield on $Y$ so that the integrand is of the form $(E d^4 x) (F_y dy) \sim E \sqrt{g_{yy}}d^4x dy$, and the $(4,0)$-form field strength $G$ enters in just such a way that we may interpret it as the conformal compensator for (modified) old-minimal supergravity \cite{Becker:2016edk}.

This interpretation is correct only in the $H_y=0$ gauge.
In general, there is a modification by an {\it a priori} unknown function ${\mathcal F}$ reducing to 1 when $H_y \to 0$ \cite{Becker:2018phr}. 
This function must be completely weightless, gauge-invariant, and Lorentz invariant and must therefore be a function of scalar combinations of the field strengths with $(\Delta, w, d, q) = (0, 0, 0, 0)$. 
From table \ref{T:Superscale} and some experimentation, we conclude that the only such invariant is
\begin{align}
h := \frac{H_y}{(\bar GG)^{1/3} F_y} 
~.
\end{align}
Requiring the Chern-Simons term to be even under parity fixes $F_y$ to be even and $H_y$ to be odd.
Therefore, ${\mathcal F}(h)={\mathcal F}(x)$ is actually a function of the square
\begin{align}
\label{E:xDef}
x := \frac{h^2}\alpha
~~~\textrm{with}~~~
\alpha = 12
~.
\end{align}
(The normalization will prove convenient later.)
This function is fixed by invariance under extended supersymmetry.
The variation of the volume action \eqref{E:vol} depends on $\mathcal F$ and its derivative in the combination 
\begin{align}
\label{E:FhatDef}
\hat {\mathcal F} := {\mathcal F} - h {\mathcal F}'
~.
\end{align}
In particular, component results will depend on this combination rather than $\mathcal F$ itself.

%%%%%%%%%%%%%%%%%%%%%%%%%%%%%%%%%%%%%%%%%%%%%%%%%%%%%%%%%%%%%%%%
\subsection{Gravitino Supercurrent}
\label{S:GrinoCurrent}
At the next order, we have the gravitino current coupling
\begin{align}
\label{E:S1}
S_1&= \frac1{\kappa^2} \int d^4x \int dy \int d^4 \theta E\, \Psi_y^\alpha J_\alpha + \mathrm{h.c.}
\end{align}
where the $(\Delta, w, d, q) = (\frac72, 1, \frac12, 0)$ current $J$ is constructed from all the fields except the gravitino. 
The fields in $S_0$ transform under $\delta_1$, and this needs to be canceled by the gravitino transformation. The linearized version of this transformation was given as \eqref{E:GravitinoLinearized1}, but we can give the complete non-linear version of it as follows: 
The $\Xi_y^\alpha$ parts of the transformation already have the correct charges as we can see in table \ref{T:Superscale}, but the $\Omega^y$ part does not. 
Firstly, $\Omega^y$ is a vector instead of a 1-form, so we will introduce a dimension-0 complex bilinear form $\mathcal G_{yy}$ to lower the index. 
Secondly, we match the conformal and $U(1)$ weights using $G$ to find
\begin{align}
\label{E:GrinoXf}
\delta_0 \Psi_y^\alpha 
	&= \Xi_y^\alpha + (\bar GG)^{-1/3} \mathcal G_{yy} \mathcal D_\alpha \Omega^y 
.	
\end{align}
In appendix \ref{S:XiInv}, we use the $\Xi$ part of this invariance to fix the gravitino current to the form
\begin{align}
\label{E:J}
J_\alpha &= - (\bar GG)^{1/3} \hat {\mathcal F} \left[ 
	\frac {3i}2 W_\alpha 
	+ \mathcal W_\alpha^y F_y
	- \frac {3i}2 G \frac{F_y}{H_y} \mathcal D_\alpha \log \hat {\mathcal F}
	\right]
~.	
\end{align}
We then impose invariance under $\Omega$ in appendix \ref{S:OmegaInv} which results in two conditions:
Firstly, it fixes the bilinear form to 
\begin{align}
\label{E:calG}
\mathcal G_{yy} = 
	{F_yF_y \over \hat {\mathcal F} }
~.
\end{align}
In particular, it is both real and symmetric.\footnote{Compare with 11D in which it has an imaginary anti-symmetric part.}
Secondly, we obtain a differential equation for ${\mathcal F}$ which can be put in the form
\begin{align}
\label{E:calFdiffEQ}
\hat {\mathcal F}( \hat {\mathcal F} - 2 x  \hat {\mathcal F}') = \hat{\mathcal F}'
~.
\end{align}
We show that this, together with the boundary condition ${\mathcal F}(0) =1$, is equivalent to the cubic equation 
\begin{align}
\label{E:FhatAlg}
\hat{\mathcal F}^{3} - 3 x \hat {\mathcal F} - 1 = 0
~.
\end{align}
This equation can be solved in terms of radicals and integrated, although we will not need a closed-form expression. 
At large values of $\hat {\mathcal F}$ we find that $\hat {\mathcal F}\approx \sqrt{3x}$. 
The discriminant $\Delta = - 16 (4 a^3 + 27 b^2) = (4\cdot 3)^3 (x^3-\frac14)$, so for large $x$, this has three real roots, two of which merge at $x=2^{-2/3}\approx 0.63$ and go off into the complex plane as $x$ decreases.
\begin{figure}[t]
\center
\scalebox{0.40}{
\includegraphics{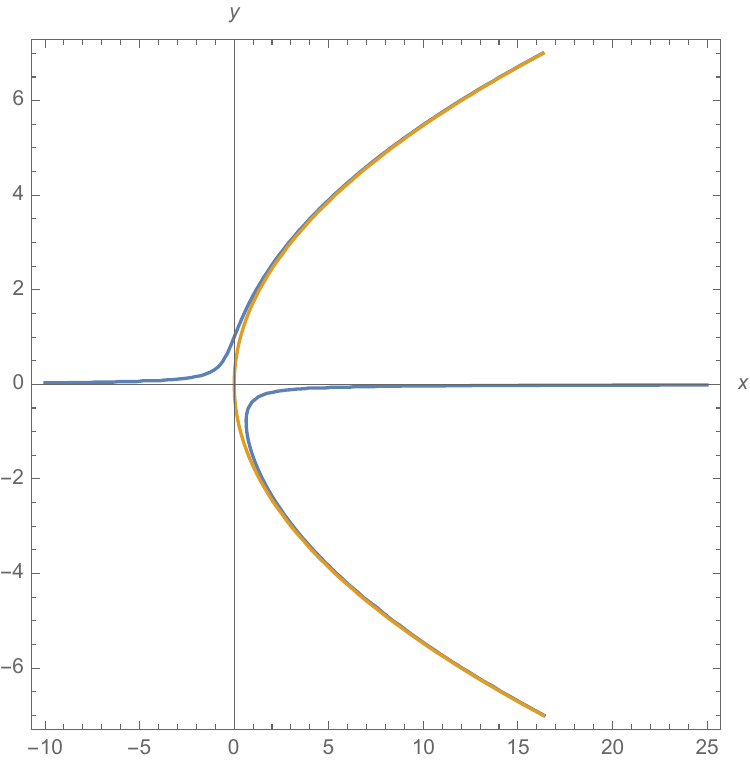}
\includegraphics{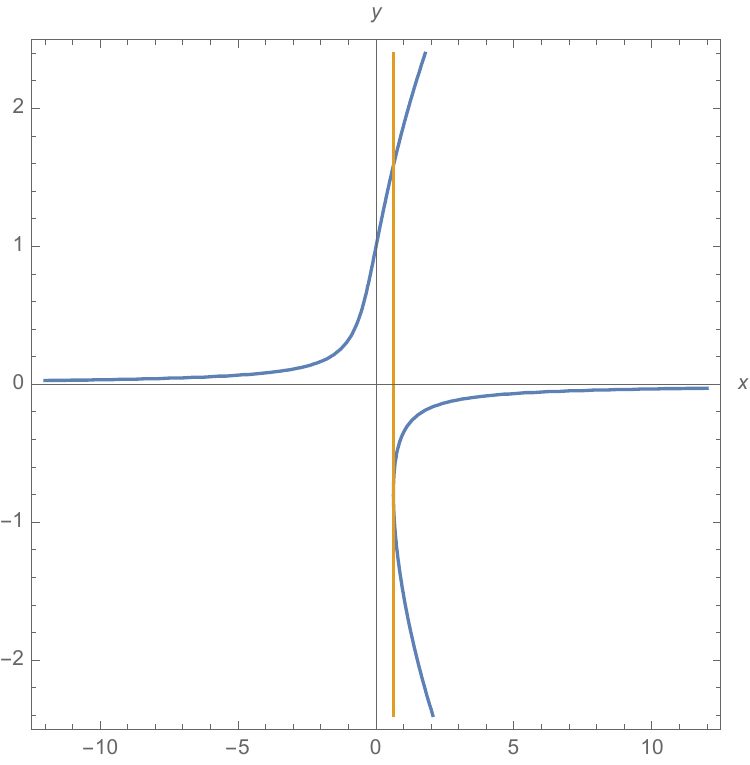}
}
\begin{caption}{
The function $\hat {\mathcal F}(x)$ has two branches interpolating between the zero function and a parabola.
On the left we plot $\hat {\mathcal F}$ and $y^2 = 3x$. On the right we give a detail of $\hat {\mathcal F}$ and the vanishing locus $x = 2^{-2/3}\approx 0.63$ of the discriminant.
}
\label{F:Fhat}
\end{caption}
\end{figure}
We plot $\hat {\mathcal F}$ as a function of $x$ in figure \ref{F:Fhat}. 
Expanding around the remaining solution, we find (for $x < 2^{-2/3}$)
\begin{align}
\label{E:Fpert}
\hat{\mathcal F} &= 1 + x -\frac {x^3}3 +\frac {x^4}3 +O(x^{6})
~~~\Rightarrow~~~
{\mathcal F} = 1 + c_1 \sqrt{x} - x + \frac {x^3}{15} -\frac {x^4}{21} +O(x^{6})
~.
\end{align}
The integration constant $c_1$ corresponds to a symmetry of the superspace
action, as shifting $c_1$ changes the action by a total derivative.
We will return to this expansion when we discuss the quadratic gravitino terms in section \ref{S:GrinoKinetic}. For reference, we also give the subleading large $x$ behavior
(for $x > 2^{-2/3}$):
\begin{align}
\label{E:FpertLarge}
\hat{\mathcal F} &= \sqrt{3x} \Big(
    1 + \frac{1}{6 \sqrt 3} \frac{1}{x^{3/2}} 
    +  O(x^{-3}) + \cdots
    \Big)
~~~\Rightarrow~~~ \notag \\
{\mathcal F} &= \sqrt{3 x} \Big(
    -\log \sqrt{x} + c_2
    + \frac{1}{18\sqrt{3}\, x^{3/2}}
    + O(x^{-3})
    \Big)
~,
\end{align}
in terms of a second integration constant $c_2$. Numerically 
matching the two solutions at $x = 2^{-2/3}$ can determine $c_2$ in terms of $c_1$ but we
won't need the explicit relation.

\subsection{Summary}
Supersymmetry has fixed that exact analytic form of the action in the first two orders of the gravitino expansion $S=S_0 + S_1+\dots$~. First, manifest supersymmetry, superscale symmetry, and abelian gauge invariance were used to construct three invariants: the volume term $S_{vol}$ \eqref{E:vol}, the Chern-Simons term $S_{CS}$ from \S{}\ref{S:CSA}, and the coupling to the gravitino current $S_1$ \eqref{E:S1}. 
Of these, only the Chern-Simons action is polynomial in the fields of the non-abelian tensor hierarchy.
The volume term is a generalization of what is usually called the K\"ahler term, and it depends on a function $\mathcal K_y =\sqrt{g_{yy}}\mathcal F(x) = F_y \mathcal F(x)$ we can loosely refer to as the K\"ahler function.\footnote{Strictly, a K\"ahler function depends only on chiral scalar fields and is defined only up to K\"ahler transformations. In our case, it depends also on 2-form superfields through the function $\mathcal F(x)$. (See {\it e.g.}\ \cite{Binetruy:2000zx} for background on such modifications in 4D, $N=1$ supergravity models.)}
The second, non-manifest supersymmetry (\ref{E:NATHgauge1}, \ref{E:GrinoXf}) was then used to fix the dependence \eqref{E:J} of the supercurrent on this function and to simultaneously determine $\mathcal F$ to be a {\em non-polynomial} function \eqref{E:Fpert} of $x$ \eqref{E:xDef}.

This is familiar from extended supersymmetric theories written in $N=1$ superspace.\footnote{For example, in \cite{Lindstrom:1983rt} Lindstr\"om and Ro\v{c}ek construct (among other things) 4D, $N=2$ tensor multiplet models by coupling $N=1$ tensor multiplets and chiral multiplets and imposing the second supersymmetry by hand as we are doing here.}
We note, though, that the non-polynomial nature of $\mathcal F(x)$ does not imply that the component action is non-polynomial in the tensor multiplets:
The variable $x$ is not a tensor under the transformations \eqref{E:NATHgauge1}; therefore, there is a Wess-Zumino gauge in which it is nilpotent, and, in such a gauge, only the first few terms of the expansion \eqref{E:Fpert} survive.
On the other hand, the non-polynomial nature of $S_{vol}$ can have important implications for higher-derivative corrections in effective actions, and we will comment on this in section \ref{S:R2}.
First however, we take an illuminating detour into the five-dimensional superspace origins of the $N=1/2$ formulation to both explain the field content we have uncovered and to identify the origin of the function $\mathcal F(x)$.

%%%%%%%%%%%%%%%%%%%%%%%%%%%%%%%%%%%%%%%%%%%%%%%%%%%%%%%%%%%%%%%%
%%%%%%%%%%%%%%%%%%%%%%%%%%%%%%%%%%%%%%%%%%%%%%%%%%%%%%%%%%%%%%%%

\section{Connecting with 5D supergravity and 5D, $N=1$ superspace}

In this section, we will identify the off-shell 5D supergravity that is most closely related to the $N=1/2$ formulation and describe two complementary perspectives on the origin of the scalar fields lying at the bottom components of $G$ and $H_y$.
Such a comparison is somewhat at odds with the general philosophy of the paper, as the point of our approach is to \emph{not} assume that the (partially) off-shell tensor calculus is known {\it a priori} or is even possible.
Nevertheless, understanding the connection is rather illuminating in explaining the presence of the various $N=1$ superfields, the roles that they are playing, and the structure of their extended supersymmetry transformations. The presentation of this section will be schematic since the details of the reduction of the component calculus is quite subtle \cite{Banerjee:2011ts} and would be even more involved in superspace.

\subsection{5D, $N=1$ supergravity and its 5D, $N=1/2$ decomposition}
\label{S:Formulations}

All formulations of 5D, $N=1$ Poincar\'e supergravity may be formulated as the standard Weyl multiplet of 5D conformal supergravity coupled to compensating multiplets. The standard Weyl multiplet is an off-shell multiplet with $32+32$ components 
\begin{align}
\label{E:sW}
sW_5 = \{e_{\bm m}{}^{\bm a} , \psi_{\bm m}{}^{\bm\alpha i} , V_{\bm m}{}^{ij}, T_{\bm {ab}}, \chi^{\bm \alpha i} , D\}
~.
\end{align} 
Its gauge sector $\{e_{\bm m}{}^{\bm a} , \psi_{\bm m}{}^{\bm \alpha i} , V_{\bm m}{}^{ij}\}$ consists of a graviton, gravitino, and an (auxiliary) $SU(2)_R$ gauge field.
In addition, it contains auxiliary covariant fields $\{T_{\bm {ab}}, \chi^{\bm \alpha i} , D\}$ consisting of a real anti-symmetric tensor, a spinor, and a real scalar. The gravitino and the auxiliary spinor are symplectic-Majorana, with spinor index $\bm\alpha=1,\cdots,8$ and $SU(2)_R$ index $i=1,2$.

As in 4D, $N=2$ \cite{deWit:1982na}, the standard Weyl multiplet must be coupled to two different types of compensators to generate a physically sensible on-shell two-derivative Poincar\'e supergravity, which in 5D will possess 48+48 degrees of freedom.
One of these compensators is always a vector multiplet, while the other may be a hypermultiplet, a non-linear multiplet, or a linear multiplet (also known as a 3-form multiplet).
The hypermultiplet requires a central charge in order to be off-shell.\footnote{One may use an off-shell hypermultiplet without a central charge in harmonic \cite{Galperin:2001uw} or projective superspace \cite{Kuzenko:2007hu}. This requires an infinite number of auxiliary fields.} The non-linear multiplet leads to the off-shell formulation of Zucker \cite{Zucker:1999ej}. The third option, a linear multiplet, turns out to be almost precisely the field content of our 5D, $N=1/2$ superspace. In particular, the 5D linear multiplet contains a 3-form $C_{\bm{mnp}}$, which explains the component fields $C_{mnp}$ and $C_{mn y}$ present in $G$ and $H_y$ respectively.

Other possible off-shell Poincar\'e supergravities may be found using the dilaton-Weyl multiplet 
\cite{Bergshoeff:2001hc} as a starting point, but let us postpone discussion of for now.

The vector and linear multiplets have the respective field content
\begin{align}
VM_5 = \{\sigma, A_{\bm m}, \psi^{\bm\alpha i}, Y^{ij}\}~, \qquad
LM_5 = \{ \ell^{ij}, C_{\bm{mnp}}, \lambda^{\bm\alpha i}, N\}~.
\end{align}
$A_{\bm{m}}$ and $C_{\bm{mnp}}$ are abelian gauge fields and $Y^{ij}$ and $N$ are (pseudo-)real auxiliaries. $\sigma$ is a real scalar and $\ell^{ij}$ is a pseudoreal scalar triplet.

Since we're just interested in understanding our superfield content, let us sketch the reduction of the 5D multiplets to 4D, $N=2$ and then to 4D, $N=1$, discarding dependence on the fifth coordinate. The reduction of these multiplets to 4D, $N=2$ (except for the linear multiplet) was worked out explicitly in \cite{Banerjee:2011ts}. Schematically, the 5D Weyl multiplet decomposes into the 4D, $N=2$ Weyl multiplet (24+24) plus a Kaluza-Klein vector multiplet (8+8). These involve the fields
\begin{align}
sW_4 = \{e_{m}{}^{a} , \psi_{m}{}^{\alpha i} , V_{m}{}^{ij}, V_m, 
    T^-_{ab}, \chi^{\alpha i} , D\}~, \qquad
VM_{KK} = \{
    X_0, A_{0 m}, \psi_0^{\alpha i}, Y_{0}^{ij}
    \}
\end{align}
where $V_m{}^{ij}$ and $V_m$ are the (auxiliary) gauge fields of the $SU(2) \times U(1)$ $R$-symmetry group in 4D. The $U(1)$ $R$-symmetry is a St\"uckelberg symmetry from the point of view of 5D, as it can be eliminated using the newly-introduced phase of $X_0 \sim e_5{}^y e^{i \varphi}$. 
The decomposition of the vector and linear multiplets is straightforward:\footnote{There are subtleties in this decomposition that we are glossing over. For example, the $\ell^{ij}$ in $LM_4$ should have Weyl weight two, but in 5D $\ell^{ij}$ has Weyl weight three, so the two are actually related by a factor of $|X_0|$. As we are only interested in organizing the component fields into superfield representations, these details can be ignored.}
\begin{align}
VM_4 = \{X, A_{m}, \psi^{\alpha i}, Y^{ij}\}~, \qquad
LM_4 = \{ \ell^{ij}, B_{mn}, C_{mnp}, \lambda^{\bm\alpha i}, N\}~.
\end{align}
The complex scalar $X$ arises from $\sigma$ and $A_y$. Note the linear multiplet $LM_4$ is a variant of the conventional 4D linear multiplet, in which one of the two real scalar auxiliaries has been replaced with the dual of a 4-form field strength $\varepsilon^{mnpq} \partial_{[m} C_{npq]}$. For further details of the dictionary, we refer to \cite{Banerjee:2011ts}.

These multiplets must now be decomposed to $N=1$. 
For simplicity, we will denote $N=1$ multiplets by their corresponding superfields.
Both vector multiplets reduce to a vector multiplet plus a chiral multiplet, while the linear multiplet decomposes into a linear multiplet plus a constrained chiral multiplet; so we write
\begin{align}
VM_{KK} \rightarrow \{\phi_0, \bar\phi_0, \mathcal V\}~, \quad
VM_4 \rightarrow \{\phi_y, \bar\phi_y, V\}~, \quad
LM_4 \rightarrow \{\Sigma_{y \alpha}, \bar \Sigma_{y \dt\alpha}, X\}~.
\end{align}
The 4D, $N=2$ Weyl multiplet is subtler. As shown \emph{e.g.} in \cite{Butter:2010sc}, it decomposes into the $N=1$ Weyl multiplet $U_{\alpha \dt\alpha}$, a conformal gravitino multiplet $\Psi_{\alpha}$, and a vector multiplet superfield $U$:
\begin{align}
sW_4 \rightarrow \{ U_{\alpha \dt\alpha}, \Psi_{\alpha}, \bar\Psi_{\dt\alpha}, U\}~. 
\end{align}
The superfield $U$ is purely auxiliary because it contains only auxiliary fields of the $N=2$ Weyl multiplet.

In comparing the above field content to 5D, $N=1/2$ superspace, we immediately identify the difference: the additional auxiliary superfield $U$ and the chiral multiplet $\phi_0$. Schematically, what occurs is that $U$ eats the chiral multiplet to become an unconstrained (non-gauge) real superfield, which can be integrated out algebraically.\footnote{More precisely, $U$ eats a linear combination of $\phi_0$ and $\phi_y$, leaving behind another linear combination, which becomes the $\phi_y$ of 5D, $N=1/2$ superspace. This is how $e_y{}^5$ ends up paired with $A_y$ in a complex scalar when, from an $N=2$ perspective, these lie in different multiplets.} This was exhibited explicitly at the linearized level in the simpler model of pure 4D, $N=2$ supergravity in \cite{Butter:2010sc}. The full non-linear description of this mechanism would presumably be quite complicated, but the underlying degrees of freedom cannot change. Therefore, we can conclude that the 5D, $N=1/2$ superspace corresponds to a partially on-shell conventional 5D supergravity with a vector and linear multiplet compensator.

We emphasize that because we have collectively integrated out $U$ and $\phi_0$, we have explicitly broken two of the $N=2$ multiplets, so the algebra of the second supersymmetry will not close off-shell.

\subsection{Spinor geometry and embedding of 5D, $N=1/2$ superspace}
\label{S:1DSpinorGeometry}

An alternative approach to understanding the spectrum of superfields we have is to consider the embedding of 5D, $N=1/2$ superspace into 5D, $N=1$ superspace. This will give us some insight into some of the structure of the $\Omega$ transformations of the extended supersymmetry.
We will only be concerned with the lowest component fields in these frames and the lowest component fields in the compensators, so all expressions should be understood to hold only on these components. 
Throughout this section (and only here), we will suppress the $|$-notation indicating projection to $\theta, \bar \theta \to 0$. 

We work in the 5D superspace conventions of \cite{Kuzenko:2005sz},
except we use $\bm\alpha$ to denote an 8-component spinor index.
In particular, we take $\varepsilon^{ij}$ and $\varepsilon_{ij}$ to obey 
$\varepsilon^{12} = \varepsilon_{21} = 1$.
Our presentation will be brief, as the analysis follows step-by-step that of reference \cite{Becker:2018phr}, to which we refer for a more leisurely exposition.

Consider reducing the 5D, $N=1$ superspace frames to those of 4D, $N=1$ superspace.
First breaking the Lorentz group to $SO(3,1)$, with
$E^{\bm \alpha}{}_i = (E^{\alpha}{}_i, - \varepsilon_{i j} E_{\dt\alpha}{}^j)$ and
$E^{\bm a} = (E^a, E^5)$, we then exhibit the leading parts of the embedding:
\begin{alignat}{2}
\label{E:ReducedFrames}
E^{\alpha}{}_i &\to \sqrt{\phi} \,\eta_{i} \,e^{i \pi/4} E^\alpha + \dots~, &\qquad
E^{\dt\alpha}{}^i &\to \sqrt{\phi} \,\bar\eta^{i} \,e^{-i \pi/4} E^{\dt \alpha} + \dots~, \notag \\
E^a &\to \phi E^a + \cdots ~, &\qquad 
E^5 &\to dy\, F_y  + \cdots~.
\end{alignat}
The spinor $\eta_{{i}}$ and its complex conjugate $\bar \eta^i$ are valued in $SU(2)$,
and they parametrize the embedding of the half of supersymmetry being kept manifest.
The phase factors $e^{\pm i\pi/4}$ are for convenience later on.
We have factored out a field $\sqrt\phi$ so that $\eta_i$ can be normalized to
$\eta_{{i}} \bar \eta^{{i}}=1$. To preserve the canonical dimension-zero torsion tensor,
$\phi$ must appear in the leading term of $E^a$ as well. We have denoted the leading
term of $E_y{}^5$ by $F_y$ for later convenience; we must still establish that this is
the correct identification.

In the 11D setting, we identified $\phi$ as the conformal compensator, which was sensible because 11D superspace has no Weyl symmetry of its own, so we had to insert that symmetry {\it \`a la} St\"uckelberg. But as we have just reviewed, 5D supergravity can be formulated \emph{already} with a Weyl symmetry when coupled to vector multiplet and a linear multiplet compensator. To avoid overcounting symmetries and to keep the situation as close to 11D as possible, let us suppose that we have put 5D supergravity on-shell and gauge-fixed as much as possible. We fix its $SU(2)_R$ symmetry to some $SO(2)_R$ subgroup by choosing a background value for the isotriplet in the linear multiplet, \textit{e.g.} $\ell^{ij} = \delta^{ij}$, and fix the Weyl symmetry by normalizing $\ell^{ij} \ell_{j k} = \delta^i{}_k$  where $\ell_{i j} = (\ell^{i j})^* = \varepsilon_{i k} \varepsilon_{j l} \ell^{k l}$.

How should we then understand our 4D, $N=1$ superconformal symmetry? The $U(1)$ $R$-symmetry group is just the residual $SO(2)$ $R$-symmetry of 5D. The Weyl symmetry corresponds to the St\"uckelberg symmetry of $\phi$, which we now identify as the conformal compensator. The fields $\eta_i$, which contain three degrees of freedom, are describing a hidden St\"uckelberg $SU(2)$ symmetry.

Let us now identify $G$ and $H_y$.
Remember that a linear multiplet is encoded in a 4-form field strength with leading term
\begin{align}
G_4 = i\, E^{\bm\alpha}_i E^{\bm \beta}_j\, E^{\bm a} E^{\bm b} (\Sigma_{\bm a \bm b})_{\bm \alpha \bm \beta}\, \ell^{i j} + \cdots
\end{align}
Upon decomposing this to 4D using the conventions of \cite{Kuzenko:2005sz}, we recover
\begin{align}
G_4 &= \phi^3\, \eta_i \eta_j \ell^{i j}\, E^\alpha E^\beta E^a E^b (\sigma_{ab})_{\alpha\beta} + \text{c.c.} + \cdots~, \cr
H_{3} = G_{3 y} &= -2 \phi^2\, F_y\, \eta_i \bar \eta^j \, \varepsilon_{j k} \ell^{k i} \, 
    E^\alpha \, E_{\dt\beta}\, E^c\, (\sigma_c)_\alpha{}^{\dt\beta} + \cdots
\end{align}
which implies that
\begin{align}
G = \phi^3\, \bar \eta^i \bar \eta^j \ell_{i j}~, \qquad
H_y = 2i \,\phi^2 \eta_i \bar \eta^j \varepsilon_{j k} \ell^{k i}\, F_y~.
\end{align}

The components $G$, $\bar G$, and $H_y$ can be interpreted as dressed versions of the
three components of the underlying linear multiplet $\ell^{ij}$. This is transparent
if we had embedded 4D superspace trivially by taking $\phi=1$ and $\eta_i = (1,0)$. Then we would
instead have left the underlying 5D Weyl and $SU(2)_R$ symmetries unfixed, so $\ell^{ij}$ would
have been a dynamical field, and $G$ and $H_y$ would have been its components.

One now easily checks that $G$ and $H_y$ satisfy
\begin{align}
\label{E:CompensatorAlg}
\bar GG  + \frac{1}{4} \left({\phi H_y\over F_y}\right)^2  =  \phi^6 
~~~\Rightarrow~~~
\frac\alpha4 x = (\bar \eta^2 \eta^2)^{-2/3} - (\bar \eta^2 \eta^2)^{1/3} 
~, \quad \eta^2 \equiv \eta_i \eta_j \ell^{i j}~.
\end{align}
We think of the conformal compensator $\phi$ now as being defined by this equation.
Note that the first equation can be interpreted as a dressed version of the gauge-fixing
condition $\frac{1}{2} \ell^{ij} \ell_{ij} = 1$.

Remember that the spinor $\eta_i$ describes three independent degrees of freedom. Its normalization condition is invariant under
\begin{align}
\delta \eta_{{i}} = - i \omega \, \eta_{{i}} - i w F_y \,\varepsilon_{{i}{j}} \, \bar \eta^{{j}}
~~~\textrm{and}~~~
\delta \bar \eta^{{i}}= i \omega \, \bar \eta^{{i}} - i \bar w\, F_y \,\varepsilon^{{i} {j}}\, \eta_{{j}}
~.
\end{align}
where $\omega$ is real and $w$ is complex. We are separating out a factor of $F_y$ for convenience.
Focusing on the complex $w$ parameter, we find induced transformations on $G$ and $H_y$:
\begin{align}
\delta G = \phi \bar w  H_y
~~~\textrm{and}~~~
\delta H_y  = - 2 \phi^{-1} F_y^2 (wG  +\bar w \bar G)
~.
\end{align}
These transformations can be identified, as in eleven dimensions \cite{Becker:2018phr}, with a piece of the $\Omega$ transformations. Letting $z := \frac i4 (\bar GG)^{1/3} \bar G^{-1} \mathcal D^2 \Omega^y|$, we find that
\begin{align}
\delta G = G (\bar GG)^{-1/3} H_y \bar z
~~~\textrm{and}~~~
\delta H_y &= - 2 (\bar GG)^{1/3} \mathcal G_{yy}  z + \mathrm{h.c.}
\end{align}
Matching the $\delta G$ transformations suggests the identification $w = \phi^{-1} \bar G (\bar GG)^{-1/3}  z$, and matching $\delta H_y$ requires (using \eqref{E:calG})
\begin{align}
\label{E:Fhat}
\hat {\mathcal F} = (\bar \eta^2 \eta^2)^{-1/3}
~.
\end{align}
Substituting this into \eqref{E:CompensatorAlg}, we find the cubic equation \eqref{E:FhatAlg} in the form
$3x  = \hat {\mathcal F}^2 - \hat {\mathcal F}^{-1}$,
provided $\alpha =12$ in \eqref{E:CompensatorAlg}.

One important feature that we glossed over was why we should identify $F_y$ as $E_y{}^5$. The key point is that the $\Omega$ transformation of $F_y$ has no $z$ piece at lowest component, and so $F_y$ cannot be associated with the embedding. Since in this subsection, we stated our starting point was on-shell gauge-fixed 5D supergravity, there is no dynamical scalar field to identify with $F_y$. The only possibility is that it should be identified with the radion $E_y{}^5$.

Now we have provided two complementary interpretations of the compensating multiplets.
From the component perspective, they may be identified as descending from a partly on-shell conventional 5D supergravity. From a superspace perspective, they correspond to St\"uckelberg fields associated with the precise embedding of 5D, $N=1/2$ superspace into 5D, $N=1$ superspace.

%%%%%%%%%%%%%%%%%%%%%%%%%%%%%%%%%%%%%%%%%%%%%%%%%%%%%%%%%%%%%%%%.%%%%%%%%%%%%%%%%%%%%%%%%%%%%%%%%%%%%%%%%%%%%%%%%%%%%%%%%%%%%%%%%
\section{Linearized Action}
\label{S:LinearizedAction}

At this point we have constructed $S_0$ and $S_1$ in the gravitino expansion \eqref{E:GrinoExpansion} to all orders in the remaining fields. 
Although it is essentially guaranteed by the component spectrum and symmetries, we would like to verify that our procedure has given the correct action by checking it explicitly. 
We could, for example, project the result to components and compare to the known component action ({\it e.g.}\ \cite{Banerjee:2011ts}).
Alternatively, we could compare the linearization of our action to a linearized superspace action known to produce the correct component result \cite{Linch:2002wg}. 
Since it is separately of interest to work out the linearized superspace action ({\it e.g.}\ for quantization) and to understand how the Higgsed superfield spectrum is reproduced correctly, we will make this comparison presently (cf.\ \S{}\ref{S:HiggsedAction}). 
But to do so, we first need the quadratic-in-gravitini action $S_2$.
For the comparison we will be making, and to show the consistency of the extended superconformal symmetries, a linearized approximation of this part of the action suffices, and we will construct it next.

We emphasize that up until now, we have kept the superspace measures for the $y$-independent 4D, $N=1$ conformal supergravity background geometry. In the remaining sections, we will work to first order in fluctuations around this background. To reduce cumbersome notation, we henceforth simply take the background to be five-dimensional Minkowski space, but one could in principle consider any background satisfying the equations of motion.

%%%%%%%%%%%%%%%%%%%%%%%%%%%%%%%%%%%%%%%%%%%%%%%%%%%%%%%%%%%%%%%%
%%%%%%%%%%%%%%%%%%%%%%%%%%%%%%%%%%%%%%%%%%%%%%%%%%%%%%%%%%%%%%%%
\subsection{Gravitino Kinetic Terms}
\label{S:GrinoKinetic}

As the volume action is a function of $x\sim H_y^2$ \eqref{E:xDef}, the ``mass'' term $(\partial_y X)^2$ appearing throughout transforms under superconformal transformations $L^\alpha$ \eqref{E:delta0X}. 
The form of this transformation is such that it can be canceled only by the $\delta_{-1}$ transformation \eqref{E:delta-1} of the combination 
%$\tfrac1{2i}\left[ {\mathcal D}^\alpha (G \Psi_\alpha) - \bar {\mathcal D}_{\dt \alpha} (\bar G\bar \Psi^{\dt \alpha})\right]$ 
${2i}B_y := {\mathcal D}^\alpha (G \Psi_{y\alpha}) - \bar {\mathcal D}_{\dt \alpha} (\bar G\bar \Psi_y^{\dt \alpha})$ 
of the gravitino superfield. 
As such, we define the combination 
\begin{align}
\label{E:Tdef}
T_y := H_y 
	+\frac1{2i}\left[ 
		{\mathcal D}^\alpha (G \Psi_{y \alpha}) - \bar {\mathcal D}_{\dt \alpha} (\bar G\bar \Psi_y^{\dt \alpha})
		\right]
\end{align}
and replace $H_y\to T_y$ in all actions constructed heretofore. 
To get the quadratic term of the gravitino, we make this replacement in the volume term and expand in $B$:
\begin{align}
\label{E:KahlerExpansion}
(\bar G G)^{1/3} F_y {\mathcal F}(h) \to  (\bar G G)^{1/3} F_y {\mathcal F}(h)
	+ B_y {\mathcal F}'(h)
	+{ {\mathcal F}''(h)B_y^2\over 2(\bar G G)^{1/3} F_y} + \dots
~,
\end{align}
The first term gives back $S_{vol}$, the second reproduces the $G \mathcal D {\mathcal F}'$ contribution of \eqref{E:J} to $S_1$ \eqref{E:S1}, and the third term gives our gravitino kinetic terms.

Similarly, expanding explicitly in the $y$-dependence of the conformal graviton, we encounter the ``mass'' term $(\partial_y U^a)^2$, which, again, can only be covariantized, provided it appears everywhere exclusively in the $(\Delta, w, d, q) = (-1, 0, 0, 1)$ combination 
\begin{align}
\label{E:Xdef}
2i X_y^{\un a} := \bar {\mathcal D}^{\dt \alpha} \Psi_y^{\alpha} + {\mathcal D}^\alpha \bar \Psi_y^{\dt \alpha} -\partial_y U^{\un a}
~.
\end{align}
%with $(\Delta, w, d, q) = (-1, 0, 0, 1)$. 
(Note that argument can be reversed to imply that the action can only depend on this particular combination of $\bar {\mathcal D}\Psi$ and its conjugate.)
As with $H_y$, this term can only enter the action as a square, which, by the charges of table \ref{T:Superscale}, must be the charge-less combination 
\begin{align}
\label{E:yDef}
y:= (\bar GG)^{1/3} { X_y^a \eta_{ab} X_y^b\over F_yF_y}
\end{align}
where $\eta_{ab}$ is the 4D Minkowski metric. 
We can fix the coefficient of this term from 5D Lorentz invariance \cite{Becker:2016edk}. In the superspace Lorentz gauge $D_\alpha U^{\un a} = 0$, the old-minimal supergravity Lagrangian reduces to $-U_a \Box U^a-\frac13 \bar GG$, and we simply pick the coefficient of $X^2$ to match this \cite{Buchbinder:2003qu, Becker:2016edk}. (The $\bar GG$ term is irrelevant to this argument.)
In the next section, we will linearize and combine the results to obtain a new formulation of the quadratic superspace action for 5D, $N=1$ supergravity.

%%%%%%%%%%%%%%%%%%%%%%%%%%%%%%%%%%%%%%%%%%%%%%%%%%%%%%%%%%%%%%%%.%%%%%%%%%%%%%%%%%%%%%%%%%%%%%%%%%%%%%%%%%%%%%%%%%%%%%%%%%%%%%%%%
\subsection{Higgsed Gravitino Action}
\label{S:HiggsedAction}

We will now linearize this formulation of 5D, $N=1$ supergravity around a Minkowski background. 
Specifically, we take $\bm X \times Y$ and replace $\bm X\to \mathbf R^{4|4}$ with flat 4D, $N=1$ superspace and $Y\to \mathbf R$.\footnote{We choose this for clarity of exposition but more generally we could replace $\bm X$ with any $y$-independent background that solves the curved 4D, $N=1$ torsion constraints.}
\begin{align}
\begin{array}{ccccccl}
{\mathcal D}_A &\to& e^{U} \vev{D_A} e^{-U} 
&~~~\textrm{with}~~~& U &=& i U^a \partial_a
\cr
X &\to& \vev{X} + X 
&~~~\textrm{with}~~~&
\vev{X} &=& \theta^2 
\cr
\Phi_y &\to& \vev{\Phi} + \Phi
&~~~\textrm{with}~~~&
\vev{\Phi} &=& i
\end{array}
\end{align}
(We use the same symbols to avoid further complicating the notation.)
The remaining fields are taken to have vanishing background values.
Note that this background breaks $q$-charge (form degree on $Y$) corresponding to fixing the internal frame $\vev{e_y{}^5} = 1$.

In the quadratic approximation, the action $S^{(2)}=\int d^5x L^{(2)}$ is given in terms of the Lagrangian that is the sum of the terms
\begin{align}
\kappa^2 L_{vol}^{(2)} &= \kappa^2 L_{omsg}[G, U^a] -2 \int d^4\theta\, \left[ {S} F_y + \mathscr L_{\mathcal V} {P}\right]
\cr
\kappa^2 L_{CS}^{(2)} &= -\frac1{4} \int d^2\theta\, \left[ 
	3W^\alpha W_\alpha - \mathcal W^\alpha \mathcal W_\alpha
	\right] +\mathrm{h.c.}
\cr
\kappa^2 L_{\Psi J}^{(2)} &= -\frac i2 \int d^4\theta\, \Psi^\alpha \left[ 
	3W_\alpha - i\mathcal W_\alpha
	\right] +\mathrm{h.c.}
\cr
\kappa^2 L_{\Psi\Psi}^{(2)} &=\int d^4\theta\, \left[ X_{ya}X_y^a + \frac14 T_y^2
	\right]
~.
\end{align}
The ingredients in the first line are as follows: $L_{omsg}[G, U^a] $ is the linearized action of old-minimal supergravity. 
It can be written in various forms, one of which is \cite{Becker:2017zwe}
\begin{align}
\label{E:LinOMSG}
L_{omsg} = \int d^4 \theta \Big\{ - U_a \Box U^a 
	+\frac18 \bar D^2 U_a D^2 U^a
	+ \frac1{48} ([D_\alpha , \bar D_{\dt \alpha}] U^{\un a})^2 
	-(\partial_a U^a)^2 
\cr	+\frac{2i}3 (G - \bar G)\partial_a U^a 
	-\frac 13\bar GG
\Big\}
~.
\end{align}
(Note that it reduces as claimed in superspace Lorentz gauge $D_\alpha U^{\un a} = 0\Rightarrow D^2 U^a = 0$.)
The remaining couplings are of the radion and KK field to scalar and pseudo-scalar combinations
\begin{align}
\begin{array}{lcl}
{S} := \tfrac12 (G+\bar G) - \tfrac14 [{D}_\alpha , \bar {D}_{\dt \alpha} ] H^{\un a} 
	&~~~\Rightarrow~~~&
	\delta {S} = \tfrac38 \left( {D}^\alpha \bar {D}^2 L_\alpha + \bar {D}_{\dt \alpha} {D}^2 \bar L^{\dt \alpha}\right)
\cr
{P} := \tfrac1{2i} (G-\bar G) - \tfrac12 \partial_{\un a} H^{\un a} 
	&~~~\Rightarrow~~~&
	\delta {P} = \tfrac i8 \left( {D}^\alpha \bar {D}^2 L_\alpha - \bar {D}_{\dt \alpha} {D}^2 \bar L^{\dt \alpha}\right)
%~.	
\end{array}
\end{align}
respectively. 
These are the real and imaginary parts of the linearized superframe determinant $G+\tfrac12 \bar {D}_{\dt \alpha} {D}_\alpha H^{\un a} \to \tfrac14 {D} \bar {D}^2 L + \tfrac12 \bar{D}  {D}^2 \bar L$.\footnote{
These combinations of the compensator of old-minimal supergravity ($n=-1/3$) transform as the compensators of new-minimal supergravity ($n=0$) and virial supergravity \cite{Nakayama:2014kua,Gates:2003cz,Buchbinder:2002gh}, respectively.
} %end footnote

To understand the physics contained in the action $S^{(2)}$, we could project to components, at least for the quadratic action. 
For the eleven-dimensional action, this was done for the linearized action in \cite{Becker:2017zwe} and for the complete bosonic scalar potential in \cite{Becker:2016edk}.
Instead, we will relate $S^{(2)}$ to the known linearized supergravity action that was formulated and projected to components in reference \cite{Linch:2002wg}.
In that formulation, the extended superconformal symmetries $\Xi$ and $\Omega$ are absent.
The associated vectors ($\mathcal V$ and $V$) and 2-form ($\Sigma$) have Higgsed the gravitino to 
\begin{align}
\label{E:HiggsedGrino}
\bm \Psi^\alpha := \Psi_y^\alpha +\Sigma_y^\alpha - {D}^\alpha \left(\mathcal V^y+iV\right)
~.
\end{align}
This Higgsed gravitino is invariant under $\Xi$ and $\Omega$ but transforms under the (linearized) non-abelian hierarchy symmetries \eqref{E:NATHgauge0} as 
\begin{align}
\label{E:deltaLLP}
\delta \bm \Psi^\alpha &= 2i {\partial_y} \bm L^\alpha
	+{D}^\alpha \bm \Omega
	+\tfrac i8 \bar {D}^2 {D}^\alpha u
\end{align}
with $\bm L^\alpha= L^\alpha -\tfrac i2 \Upsilon^\alpha$ and $\bm \Omega = -\frac{1}{2} \Lambda$.
This is the transformation rule of \cite{Linch:2002wg,Gates:2003qi}; in particular, the new $\bm \Omega$ is chiral and $\Xi$ has been replaced by $\bar D^2 Du$.
%\footnote{To compare, one needs to rescale $\Psi$ by a factor of $2i$---that is, to set $\hat \Psi = -\frac i2 \bm \Psi$.} 
Indeed, rewriting $S^{(2)}$ in terms of the Higgsed gravitino, we find after many cancellations that the Lagrangian collapses to
\begin{align}
\label{E:LLP}
\kappa^2 L^{(2)} = \kappa^2 L_{omsg}^{(2)} 
	-\frac12  \int d^4\theta\, \left[
		\bm X_{\un a}^2 
		-\frac12 \bm T^2 
		- 2i {S} (\Phi-\bar \Phi)
	\right]
~,	
\end{align}
where now $\bm X$ and $\bm T$ are defined in terms of the Higgsed gravitino \eqref{E:HiggsedGrino}.
This is the form of the Lagrangian density found in \cite{Linch:2002wg}. 
Although the cancellations needed to recover this form are non-trivial, the result was guaranteed since the gauge transformation \eqref{E:deltaLLP} of that reference was reproduced by the combination \eqref{E:HiggsedGrino}.

\subsection{An alternative linearization}
\label{S:NewMinimal}
While we have successfully recovered the Lagrangian of \cite{Linch:2002wg}, there is a curious point. The construction of \cite{Linch:2002wg} was built around 4D old minimal supergravity as a starting point. It is well-known that there is an alternative -- 4D new minimal supergravity -- and one might have expected to be able to find a linearization of 5D, $N=1$ supergravity built upon that. This is all the more pressing, because linearized 4D, $N=2$ supergravity (to which 5D, $N=1$ supergravity can be dimensionally reduced and then truncated) leads to a continuum of actions when rewritten in $N=1$ language \cite{Butter:2010sc}. This continuum is related to the vacuum expectation value for the linear multiplet $\ell^{ij}$ with one particular limit corresponding to old minimal supergravity and the other to new minimal. Equivalently, the two versions are related to different ways of embedding 4D, $N=1$ into 4D, $N=2$. Based on the discussion in section \ref{S:Formulations}, the same should hold here.
 
Implicit in the analysis in sections \ref{S:GrinoKinetic} and \ref{S:HiggsedAction} was the idea that $G$ takes a VEV, but $H_y$ does not. This corresponds exactly to the choice for $\ell^{ij}$ that led to old minimal supergravity in \cite{Butter:2010sc}. Instead, we shall now require the reverse: We let $H_y$ take a VEV and $G$ not.
For the original superspace action involving $\mathcal F(x)$, this corresponds to expanding around $x =\infty$. From the series expansion \eqref{E:FpertLarge}, one can show that
\begin{align}
\kappa^2 L_{vol} &= -3 \int d^4\theta\, (G \bar G)^{1/3} F_y \mathcal F(x) \notag\\
    &=
    \int d^4\theta \Big(\frac{3}{2} H_y \log (H_y / F_y) 
    -2 \,G \bar G \frac{F_y^3}{H_y^2} + \cdots\Big)
\end{align}
where the infinite series of suppressed terms are higher order in $G$.
The leading term resembles the Lagrangian for new minimal supergravity where
$H_y$ is the tensor multiplet compensator.
This is apparent from the $L_\alpha$
transformation \eqref{E:LalphaTrafo}, which identifies $\Sigma_y^\alpha$ as the
tensor multiplet compensator in this case.

Let us now linearize.
In section \ref{S:HiggsedAction}, we assumed $\vev{X}=\theta^2$ and $\vev{\Phi}=i$,
so it followed that $\vev{G}=1$ and $\vev{F_y} = 1$. From \eqref{E:CompensatorAlg},
the equivalent choice here, related to the prior one by an $SU(2)_R$ rotation,
should be $\vev{H_y}=2$ and $\vev{F_y}=1$. Linearizing about this background,
we find
\begin{align}
\kappa^2 \mathcal L_{vol} = 
    \frac{3}{2} \Big(F - \tfrac{1}{2} H\Big)^2
    - \frac{1}{2} \widehat G \widehat {\bar G}
~.    
\end{align}
The linearized curvatures $F$ and $H$ are given simply by \eqref{E:NATHFS1} and \eqref{E:NATHFS3},
%\begin{align}
%F = \frac{1}{2i} (\Phi - {\bar\Phi}) - \partial V ~, \qquad
%% 
%H = \frac{1}{2i} (D^\alpha \Sigma_\alpha - \bar D_{\dot\alpha} \bar \Sigma^{\dot\alpha}) - \partial X ~,
%\end{align}
but the linearized chiral curvature $\widehat G$ is slightly more complicated:
\begin{align}
\widehat G &:= -\frac{1}{4} \bar D^2 (X + 2 i \mathcal V ) = G - \frac{i}{2} \bar D^2 \mathcal V~.
\end{align}
The Chern-Simons term is unchanged, but the gravitino supercurrent is a bit more involved,
as one must consistently work in the $x\rightarrow \infty$ limit. 

Including also the gravitational prepotential $U^a$, one finds the quadratic action $S^{(2)}=\int d^5x L^{(2)}$ has a Lagrangian given as the sum of terms
\begin{align}
\kappa^2 L_{vol}^{(2)} &=  \int d^4\theta\, \Big[
    - U_a \Box U^a 
    +\frac18 \bar D^2 U_a D^2 U^a
    + \frac1{16} ([D_\alpha , \bar D_{\dt \alpha}] U^{\un a})^2 
    -(\partial_a U^a)^2 
    \cr & \quad
    + \frac{1}{4} H_y [D_\alpha, \bar D_{\dot\alpha}] U^{\ul a} 
    +2 \,\mathcal V \,\partial_y \partial_{a} U^{a} 
    + \frac{3}{2} \Big(F_y - \tfrac{1}{2} H_y \Big)^2
    - \frac{1}{2} \widehat G \widehat {\bar G} \Big]~, \cr
\kappa^2 L_{CS}^{(2)} &= -\frac{1}{4} \int d^2\theta\, \Big(
        3 W^\alpha W_\alpha
        - \mathcal W^\alpha \mathcal W_\alpha\Big)
    + \text{h.c.}~, \cr
\kappa^2 L_{\Psi J}^{(2)} &= -\frac{i}{2} \int d^4\theta\, \Psi^\alpha \Big(
    3 W_\alpha + i \mathcal W_\alpha
    + \frac{1}{2} D_\alpha \widehat G
    \Big)
    + \text{h.c.}~, \cr
\kappa^2 L^{(2)}_{\Psi^2} &= \int d^4\theta  \,X_{y a} X_y{}^a~.
\end{align}

The first term $L^{(2)}_{vol}$ includes the Lagrangian for new minimal supergravity involving the supergravity prepotential $U_a$ and tensor multiplet $H_y$, albeit with a slightly unusual normalization of the tensor multiplet due to the non-standard background value $\vev{H_y}=2$. The additional pieces, including $\mathcal V$ and the cross-terms between $F_y$ and $H_y$ are required for gauge-invariance.

As before, we Higgs the gravitino, but this time only to remove the $\Omega$ transformation
\begin{align}
\bm \Psi^\alpha := \Psi_y^\alpha - {D}^\alpha \left(\mathcal V^y + i V\right)
~.
\end{align}
We cannot remove the $\Xi_y^\alpha$ transformation as before, because $\Sigma_y^\alpha$ no longer transforms under this symmetry (since $\vev{G} = 0$). We similarly eliminate the $\Omega$
transformation of $X$ by defining
\begin{align}
\bm X: = X - 2 V~, \qquad
\bm H_y := \frac{1}{2i} (D^\alpha \Sigma_{y \alpha} - \bar D_{\dt\alpha} \bar \Sigma_{y}^{\dt\alpha})
    - \partial_y \bm X~, \qquad
\bm G := -\frac{1}{4} \bar D^2 \bm X~.
\end{align}
In terms of these quantities, the action reduces to
\begin{align}
\kappa^2 L^{(2)} &= \kappa^2 L^{(2)}_{nmsg}
    + \int d^4\theta\, \Big[
    -\frac{1}{2} \bm X_{\ul a}^2
    -\frac{1}{2} \bm G \bar {\bm G}
    + \frac{3}{4} \Phi_y \bar\Phi_y
    + \frac{3i}{4} \bm X \partial_y (\Phi_y-\bar\Phi_y)
    \notag \\ & \qquad\qquad\qquad\qquad\qquad\qquad\qquad\qquad\qquad
    - \frac{1}{4} ( i \bm \Psi^\alpha D_\alpha \bm G + \text{h.c.})
    \Big]
\end{align}
where $L^{(2)}_{nmsg}$ is the new minimal supergravity action built from $U_a$ and $\bm H_y$.
This action is invariant under the gauge transformations
\begin{align}
\delta \bm\Psi_{\alpha} &= \Xi_{y\alpha} - \frac{1}{2} D_\alpha \Lambda + 2i \partial_y L_\alpha~, \cr
\delta \bm X &= \frac{1}{2i} (D^\alpha \Upsilon_\alpha - \bar D_{\dt\alpha} \bar \Upsilon^{\dt \alpha})
    + i (\Lambda - \bar \Lambda)~, \cr
\delta \Phi_y &= \partial_y \Lambda~, \cr
\delta \Sigma_{y\alpha} &= -\frac{1}{4} \bar D^2 D_\alpha u + \partial_y \Upsilon_\alpha
    + i \bar D^2 L_\alpha~, \cr
\delta U_{\alpha \dt\alpha} &= \bar D_{\dt\alpha} L_\alpha - D_\alpha \bar L_{\dt\alpha}~.
\end{align}

%%%%%%%%%%%%%%%%%%%%%%%%%%%%%%%%%%%%%%%%%%%%%%%%%%%%%%%%%%%%%%%%%%%%%%%%%%%%%%%%%%%%%%%%%%%%%%%%%%%%%%%%%%%%%%%%%%%%%%%%%%%%%%%%
\section{Application: Gravitational Chern-Simons Term}
\label{S:R2}
At this point we have recast 5D, $N=1$ supergravity in terms of superfields on what could be described as 5D, $N=1/2$ superspace. 
We then verified that it reduces correctly in the linearized limit.
This essentially un-gauge-fixes the formalism of \cite{Linch:2002wg} and non-linearizes it.
In this section, we will demonstrate how it can be used to study higher-derivative corrections to the supergravity action by deriving the major part of the superspace expression for the gravitational Chern-Simons term $\sim \int A \wedge \mathrm{tr} (R \wedge R)$. 
Here $A$ stands for the graviphoton and $R$ for the curvature 2-form, so this is a purely gravitational version of the mixed gauge/gravitational Chern-Simons term. 
%\added{We will essentially be working to linearized order in $A$ and $R$, so the result will be valid to cubic order in gravitational fields.}
%\added
{We will be working with linearized invariants, but the output of the procedure is in terms of field strengths, curvatures, and torsions. 
As such, we expect the formal result to be valid beyond this order.
(The understanding of such terms|should they be present|will have to await the completion of the Kaluza-Klein super-geometry alluded to in \S{}\ref{E:GrinoPunt}.)}

The supersymmetric completion of $\int A \wedge \mathrm{tr} (R \wedge R)$ was worked out in components in \cite{Hanaki:2006pj}, and embedded into conformal superspace in \cite{Butter:2014xxa}.
%\footnote{
(Strictly speaking, what was constructed there is the mixed gauge/gravitational Chern-Simons Lagrangian in which the field $A$ is part of a generic matter vector multiplet.
In \S{}\ref{S:Discussion}, we will comment more extensively on this point.)
%}
In principle, it should be possible to reduce the latter to $N=1/2$ superspace by gauge fixing to a finite number of non-auxiliary superfields and performing the harmonic integrals. 
In practice, however, it seems much easier to construct the required composite field strengths directly, as even at the component level, the reduction from 5D to 4D is non-trivial \cite{Banerjee:2011ts}.

To demonstrate this approach, we will construct a specific class of terms that contribute to
the Chern-Simons term. Under the decomposition of $SO(4,1) \rightarrow SO(3,1)$,
\begin{align}
A \wedge \mathrm{tr} (R \wedge R) \rightarrow A \wedge R^{ab} \wedge R_{ab}
    + 2 A \wedge R^{a\,5} \wedge R_{a\,5}~.
\end{align}
From the 4D point of view, these correspond to two different classes of terms, as each is separately invariant under $SO(3,1)$ and the gauge transformation of $A$. 
In this section, we consider the $N=1$ supersymmetrization of the first term, since this is the most non-trivial part to covariantize.
(The second term can be integrated by parts (in the linearized approximation) to $F \wedge \omega^{a \, 5} \wedge R_{a \,5}$ and there seems to be no obstruction to realizing this as a full superspace integral involving covariant 4D quantities.)

To this end, we start with the superspace form of the gravitational Chern-Simons action $S_{gCS} = \int d^4 x \int_Y L_{gCS}$. By gauge invariance, supersymmetry, and so forth, this action must be identical in structure to that of the 2-derivative Chern-Simons Lagrangian \eqref{E:CSLagrangian}, but now with the composite 4-form $\{\mathbb G,\mathbb H_y\}$ constructed in terms of the 4D, $N=1$ curvature tensor superfields and their analogs with one leg along $Y$.
The reduced field strength of the 4D, $N=1$ Weyl tensor is the chiral field $W_{\alpha \beta \gamma}$, so we expect $\mathbb G = W^{\alpha \beta \gamma} W_{\alpha \beta \gamma} +\cdots$. 
To be gauge invariant, the five-dimensional 4-form represented by $\{\mathbb G,\mathbb H_y\}$ must be gauge invariant and closed. 
That is, we seek a reduced 3-form field strength $\mathbb H_y$ quadratic in 5D supergravity invariants that satisfies \eqref{E:NATHBIH}.

%%%%%%%%%%%%%%%%%%%%%%%%%%%%%%%%%%%%%%%%%%%%%%%%%%%%%%%%%%%%%%%%
\subsection{Linearized Field Strengths}
The linearized invariants of 5D supergravity in $N=1/2$ superspace were studied in \cite{Gates:2003qi}. 
A generating set is given as follows:\footnote{The normalizations of the first three superfields differ from \cite{Wess:1992cp}, with $G_{\ul a}$ and $R$ here equal to twice the same quantities in \cite{Wess:1992cp}, whereas $W_{\alpha\beta\gamma}$ differs by a factor of $-2$.}

\begin{subequations}
\label{E:FS}
\begin{align}
W_{\alpha \beta \gamma} &= \tfrac i{8} \bar D^2 D_{(\alpha} \partial_\beta{}^{\dt \gamma} U_{\gamma)\dt \gamma}
\\
G_{\un a} &= \tfrac18 D^\beta \bar D^2 D_\beta U_{\un a} 
	- \tfrac 1{24} [D, \bar D]_{\un a} [D, \bar D]^{\un b} U_{\un b} 
	- \tfrac 1{2} \partial_{\un a} \partial^{\un b} U_{\un b} 
	-\tfrac i3 \partial_{\un a} (G-\bar G)
\\
R &=  -\tfrac 1{12} \bar D^2 \left( \bar G + i \partial_{\un a} U^{\un a}\right)
\\
F'_{\alpha \beta\, y} &= 2\bar D^{\dt \beta} D_{(\alpha} X_{\beta) \dt \beta\, y}
\\
\lambda_y^\alpha &= 2 \bar D_{\dt \alpha} X_y^{\un a} - D^\alpha T_y
~.
\end{align}
\end{subequations}
The first three invariants are the irreducible parts of the linearized super-Riemann tensor \cite{Gates:1983nr,Wess:1992cp,Buchbinder:1998qv}:
$W_{\alpha \beta \gamma} $ contains the Weyl tensor, $G_a$ contains the traceless part of the Ricci tensor, and $R$ contains the curvature scalar. 
In this linearized form, it is easy to check that they satisfy the Bianchi identities 
\begin{subequations}
\label{E:BI4D}
\begin{align}
D^\gamma W_{\alpha \beta \gamma} &= -\tfrac12 \bar D^{\dt \beta} D_{(\alpha} G_{\beta)\dt \beta}
\\
\bar D^{\dt \alpha} G_{\un a} = D_\alpha R 
~~~&\Rightarrow~~~
\bar D^2 G_{\un a} = -4i \partial_{\un a} R
~.
\end{align}
\end{subequations}
In four dimensions, the equation of motion of the prepotential $U^a$ is $G_a=0$ and that of the scale compensator is $(R+\bar R)=0$, so on shell the only surviving components are the $\theta\to 0$ projections of $W_{\alpha \beta \gamma}$ and $D_{(\delta} W_{\alpha \beta \gamma)}$ corresponding to the field strengths of the helicity-$\tfrac32$ gravitino and the helicity-2 graviton, respectively.
When lifted to five dimensions, both equations of motion receive corrections so this conclusion is modified.

The remaining invariants are the gravitino multiplet's field strength $F'_{\alpha \beta \,y}$ and its equation of motion $\lambda_y^\alpha$. 
Other forms for the gravitino field strength are related to linear combinations of this one and $D^{(\alpha} \lambda^{\beta)}_y$.
The advantage of this combination is that it is complex-linear $\bar D^2 F'_{\alpha \beta \,y} =0$.
It is not difficult to verify that these satisfy\footnote{This corrects some unfortunate typographical errors in \cite{Gates:2003qi}. (We also use weighted index symmetrizations $T_{((\alpha \beta\cdots ))} = T_{(\alpha \beta\cdots )}$ instead of the unweighted ones of that reference.)}
\begin{subequations}
\label{E:BI5D}
\begin{align}
\bar D^2 F'_{\alpha \beta \,y} &=0
\\
\label{E:D2F'}
D^\beta \bar D_{\dt \alpha} F'_{\alpha \beta y} 
&=-8 \partial_y G_{\un a} 
	+\tfrac1{6}\left( D_\alpha \bar D_{\dt \alpha} +2 \bar D_{\dt \alpha} D_\alpha \right)\left[ 2 D^\alpha \lambda_{\alpha y} 
		- \bar D_{\dt \alpha} \bar\lambda_y^{\dt \alpha}\right]
\\
\label{E:D3F'}
\bar D^2 D_{(\alpha} F'_{\beta\gamma) y} &= 32 \partial_y W_{\alpha \beta \gamma} 
\\
\bar D^2 D^{\alpha} \lambda_{\alpha y} &= 48 \partial_y R
~.
\end{align}
\end{subequations}
We note for use below that the complicated identity \eqref{E:D2F'} can also be written as 
\begin{align}
\label{E:D2F'alt}
D^\beta \bar D_{\dt \alpha} F'_{\alpha \beta y} 
+ \tfrac14 D_\alpha \bar D^2 \bar \lambda_{\dt \alpha y}
&=-8 \partial_y G_{\un a} 
	+\tfrac13 D_\alpha \bar D_{\dt \alpha} \left[ D^\beta \lambda_{\beta y} -2 \bar D_{\dt \beta} \bar\lambda_y^{\dt \beta} \right]
\cr&\hspace{20mm}		
	- \tfrac13 \bar D_{\dt \alpha} D_\alpha \left[ \bar D_{\dt \beta} \bar\lambda_y^{\dt \beta}- 2D^\beta\lambda_{\beta y} \right]
~.
\end{align}
In this form, the right-hand side is manifestly real. 
Explicitly, 
\begin{align}
\label{E:D2F'real}
D^\beta \bar D_{\dt \alpha} F'_{\alpha \beta y} + 
\bar D^{\dt \beta} D_{\alpha} \bar F'_{\dt \alpha \dt \beta y}
&=- \tfrac14D_\alpha \bar D^2 \bar \lambda_{\dt \alpha y} 
+ \tfrac14\bar D_{\dt \alpha} D^2  \lambda_{\alpha y} 
~.
\end{align}

%%%%%%%%%%%%%%%%%%%%%%%%%%%%%%%%%%%%%%%%%%%%%%%%%%%%%%%%%%%%%%%%
\subsection{Curvature 4-form}
Returning to the Chern-Simons form, we seek $\mathbb H_y$ such that $\bar D^2 \mathbb H_y=4 \partial_y \mathbb G$ where $\mathbb G$ contains the term $W_{\alpha \beta \gamma}^2$. 
Then, taking $\mathbb H_y= -\tfrac14 D^\gamma F_y^{\prime \alpha \beta} W_{\alpha \beta \gamma} +\cdots$ would give the relation we want by \eqref{E:D3F'}, were it not for the fact that $\mathbb H_y$ is required to be real. 
Since this term is not, we get additional terms from its conjugate that cannot be written as $\partial_y\mathbb G$ for any chiral $\mathbb G$. 
To cancel these terms, we need to add a bilinear, specifically 
$-\tfrac 18 \bar F_y^{\prime \dt\alpha \dt \beta}D^{\alpha} \bar D_{\dt \beta} G_{\un a}$. 
But again this is not real so we must add more terms to cancel those coming from the conjugate. 
The process terminates because of the structure of the invariant: $\mathbb H_y$ is of the form $\Theta_L \otimes \Theta_R$ where $\Theta_L$ is one of $F'_{\alpha \beta y}$ or $\lambda_{\alpha y}$ or derivatives thereof and $\Theta_R$ is one of $W_{\alpha \beta \gamma}$, $G_a$, or $R$ and derivatives thereof. 
The terms can be organized in order of non-increasing helicity of the reduced field strengths. 
At each step of the process of the computation of $\bar D^2 \mathbb H_y$, the helicity decreases. 
The process terminates at the term $\tfrac 1{32} D^2 \lambda_y^\alpha D_\alpha R$, because the complex conjugate of this is linear ({\it i.e.}\ annihilated by $\bar D^2$).
To carry this out, we repeatedly use the Bianchi identities \eqref{E:BI4D} and \eqref{E:BI5D}. 
In doing so, we find 
\begin{subequations}
\label{E:RR}
\begin{align}
\label{E:RR1}
\mathbb G &= W^{\alpha \beta \gamma} W_{\alpha \beta \gamma} 
	- \tfrac 14 \bar D^2 (G^a G_a)
\\
\mathbb H_y &= -\tfrac14 D^\gamma F_y^{\prime \alpha \beta} W_{\alpha \beta \gamma}
	+\tfrac 18 F_y^{\prime \alpha \beta}\bar D^{\dt \alpha}  D_\beta G_{\un a}
	+\tfrac 18 \bar D^{\dt \alpha} F_y^{\prime \alpha \beta}D_\beta G_{\un a} 
\cr	&+\tfrac 1{24} \left( D_\alpha \bar D_{\dt \alpha} + 2 \bar D_{\dt \alpha} D_\alpha \right) 
D^\beta \lambda_{\beta y} \, G^{\un a} 
	+\tfrac 1{32} D^2 \lambda_y^\alpha D_\alpha R
	+ \mathrm{h.c.}
\end{align}
\end{subequations}

We have presented the result in the form in which we originally found it, but there are alternatives. 
For example, using \eqref{E:D2F'alt}, we can put $\mathbb H$ in the form
\begin{align}
\mathbb H_y &= -\tfrac14 D^\gamma F_y^{\prime \alpha \beta} W_{\alpha \beta \gamma}
	+\tfrac 18 F_y^{\prime \alpha \beta}\bar D^{\dt \alpha}  D_\beta G_{\un a}
	+\tfrac 18 \bar D^{\dt \alpha} F_y^{\prime \alpha \beta}D_\beta G_{\un a} 
	+\tfrac 1{32} D^2 \lambda_y^\alpha D_\alpha R
	+ \mathrm{h.c.}
\cr	&	- \partial_y(G_a^2)
	+\tfrac 1{8} \left( D^\beta \bar D_{\dt \alpha} F'_{\alpha \beta y} 
	+\tfrac 14 D_\alpha \bar D^2 \bar \lambda_{\dt \alpha y} \right) G^{\un a}
~.
\end{align}
The coefficient of the $\partial_y(G_a^2)$ term is just so that it forms a cocycle with the $\bar D^2 (G_a^2)$ term in \eqref{E:RR1}.
This implies that if we subtract the former from $\mathbb H_y$, the latter will be removed from $\mathbb G$. Recalling \eqref{E:D2F'real}, the remaining term is real, so we may redefine our form to 
\begin{subequations}
\label{E:RR'}
\begin{align}
\mathbb G' &= W^{\alpha \beta \gamma} W_{\alpha \beta \gamma} 
\\
\mathbb H_y' &= -\tfrac14 D^\gamma F_y^{\prime \alpha \beta} W_{\alpha \beta \gamma}
	+\tfrac 18 F_y^{\prime \alpha \beta}\bar D^{\dt \alpha}  D_\beta G_{\un a}
	+\tfrac 18 \bar D^{\dt \alpha} F_y^{\prime \alpha \beta}D_\beta G_{\un a} 
\cr	&	+\tfrac 1{16} \left( D^\beta \bar D_{\dt \alpha} F'_{\alpha \beta y} 
	+\tfrac 14 D_\alpha \bar D^2 \bar \lambda_{\dt \alpha y} \right) G^{\un a}
	+\tfrac 1{32} D^2 \lambda_y^\alpha D_\alpha R
	+ \mathrm{h.c.}
\end{align}
\end{subequations}
In this form, $\mathbb H_y'$ is manifestly real again. 
Another form in which it manifestly satisfies the descent relation \eqref{E:NATHBIH} is
\begin{align}
\mathbb H_y' &=  -\tfrac14 D^\gamma  F_y^{\prime \alpha \beta} W_{\alpha \beta \gamma} 
	+\bar D_{\dt \alpha} \bar {\mathbb Z}_y^{\dt \alpha} 
~~~~~\textrm{with}	
\\
\bar {\mathbb Z}_y^{\dt \alpha}  &= 
-\tfrac14 \bar F'_{\dt \beta \dt \gamma y} \bar W^{\dt \alpha \dt\beta \dt\gamma}
-\tfrac 18 F'_{\alpha \beta y}D^\beta G^{\un a} 
+\tfrac 1{8} D^{\beta} \bar F_y^{\prime \dt \alpha \dt \beta}  G_{\un b}
	-\tfrac 1{32} D^2 \lambda_{\alpha y} G^{\un a} 
	+\tfrac 1{32} \bar D^2 \bar \lambda_y^{\dt \alpha} \bar R
~.
\nonumber
\end{align}

The appearance of a trivial cocycle in \eqref{E:RR} is characteristic of dimensional reduction in superspace \cite{Kuzenko:2005sz}. 
This suggests that the term we moved is necessary for five-dimensional Lorentz invariance. 
(A perhaps related observation is that moving this term violates the $\Theta_L\otimes \Theta_R$ rule as it requires $\Theta_L\sim \partial_y G$, which is not one of $F'_y$ or $\lambda_y$.)
Reintroducing the Ricci-squared and scalar curvature-squared terms, there is {\it a priori} a 2-parameter family of invariants $(\mathbb G^{(a,b)}, \mathbb H^{(a,b)})$ with 
\begin{align}
\mathbb G^{(a,b)} &= \mathbb G'
	- \tfrac 14 \bar D^2 (a G^a G_a + 2b \bar RR)
\cr	
\mathbb H_y^{(a,b)} &= \mathbb H_y'
	- \partial_y (a G^a G_a + 2b \bar RR)
~.
\end{align}
The difference between an action constructed from the primed invariants and the $(a,b)$ invariants
can be rewritten as a covariant term ({\it i.e.} not Chern-Simons), as a superspace integral of
$F_y (a G^a G_a + 2b \bar R R)$.

One cannot fix $a$ and $b$ by a purely $N=1$ argument. One choice, involving $a=b=1$ reproduces the 4D Gauss-Bonnet invariant (cf.\ {\it e.g.}\ \S{}5.6.5 of \cite{Buchbinder:1998qv}) multiplying $F_y$, but this is not the right answer in our case. Instead one can fix $a$ and $b$ by matching to 4D truncation of \cite{Hanaki:2006pj}, which suggests that the combination above must be chosen to reproduce
$R^{abcd} R_{abcd} - \tfrac{4}{3} R_{ab} R^{ab} + \tfrac{1}{6} R^2$. This corresponds to setting $a=-1/3$ and $b=-1/12$.

We emphasize that this is only that part of the full gravitational Chern-Simons invariant that is the most non-trivial to construct in this partially on-shell superspace.
Additional covariant D-terms including, for example, $F_y F'_{\alpha \beta y} F'^{\alpha \beta y}$ should be included to recover the complete 5D, $N=1$ invariant.
%one should construct what is likely a covariant full superspace integral, at leading order including a term like $F_y F'_{\alpha \beta y} F'^{\alpha \beta y}$. 
Finding this would require either matching the most general superspace expression to components or analyzing 5D Lorentz invariance at the 4D superfield level.
%, both of which would be more involved.

%%%%%%%%%%%%%%%%%%%%%%%%%%%%%%%%%%%%%%%%%%%%%%%%%%%%%%%%%%%%%%%%
\subsection{Elaboration on the Relation to Previous Work}
\label{S:Discussion}
We now pause to discuss the relation of our gravitational Chern-Simons invariant to known invariants. 
To do so, we recall from section \ref{S:Formulations}, that there are two formulations of conformal supergravity in five dimensions referred to as ``standard Weyl" \eqref{E:sW} and ``dilaton-Weyl'' \eqref{E:dW}. What they have in common is the graviton, gravitino, and $SU(2)$ gauge field $\{e_{\bm m}{}^{\bm a} , \psi_{\bm m}{}^{\bm\alpha i} , V_{\bm m}{}^{ij}\}$. The standard Weyl multiplet further includes a real 2-form, a spin-1/2 field, and a real scalar field $\{T_{\bm {ab}}, \chi^{\bm \alpha i}, D\}$ which are all auxiliary. 
By contrast, the dilaton-Weyl multiplet \cite{Bergshoeff:2001hc, Fujita:2001kv} (see also \cite{Nishino:2000cz}) further includes a scalar, a gauge 1-form, a gauge 2-form, and a spin-1/2 field $\{\varphi, A_{\bm m}, B_{\bm {mn}}, \lambda^{\bm \alpha i}\}$. 
In other words, for the standard Weyl multiplet \eqref{E:sW}, one may substitute the dilaton-Weyl multiplet
\begin{align}
\label{E:dW}
dW_5 = \{e_{\bm m}{}^{\bm a} , \psi_{\bm m}{}^{\bm \alpha i} , V_{\bm m}{}^{ij}, 
	\varphi , A_{\bm m}, B_{\bm {mn} }, \lambda^{\alpha i} \}
~.
\end{align}
It can be obtained from the standard Weyl multiplet by coupling conformally to a vector multiplet, computing the equations of motion of the latter, and defining the auxiliary fields $\{T_{\bm {ab}}, \chi^{\bm \alpha i} , D\}$ in terms of the physical fields of the vector multiplet. (In the process, we solve a Maxwell equation of the form $\partial^{\bm m} (T_{\bm {mn}} - \partial_{ [\bm m} A_{\bm n]}) +\cdots\sim 0$, so that a 2-form $B_{\bm {mn}}$ appears as $T_{\bm {mn}}\sim \partial_{ [\bm m} A_{\bm n]} + \epsilon_{\bm {mn}}{}^{\bm{pqr}} \partial_{\bm p} B_{\bm {qr}}$.)
While this adds the $8+8$ components of a vector multiplet, the equations of motion remove the same number of degrees of freedom, returning us to a $(32+32)$-component formulation.
Any action involving the standard Weyl multiplet can be replaced with the dilaton-Weyl multiplet by substituting the auxiliary fields $\{T_{\bm {ab}}, \chi^{\bm \alpha i}, D\}$ with appropriate combinations of field strengths of $\{\varphi, A_{\bm m}, B_{\bm {mn}}, \lambda^{\bm \alpha i}\}$ \cite{Bergshoeff:2001hc}. The converse is not true.

The invariant constructed in \cite{Hanaki:2006pj} is the mixed gauge/gravitational Chern-Simons action $\sim \int A \wedge \mathrm{tr} (R \wedge R)$ in which the field $A$ is part of a generic matter vector multiplet.  It was constructed in the Poincar\'e supergravity in the standard Weyl formulation analogous to ours. On the other hand, a purely gravitational Chern-Simons action in which $A$ is the {\em graviphoton} was constructed in the dilaton-Weyl formulation in \cite{Bergshoeff:2011xn}.
(Details on the difference and further developments can be found in \cite{Ozkan:2013uk, Ozkan:2013nwa}).
What we are sketching in superspace in this paper is the purely gravitational invariant of the latter \cite{Bergshoeff:2011xn} in the supergravity formulation of the former \cite{Hanaki:2006pj}.

Two natural questions arise. The first is whether we can construct the mixed gauge/gravitational invariant of \cite{Hanaki:2006pj}. We expect this to be trivial, provided we introduce additional matter vector multiplets as in \cite{Paccetti:2004ri}.
A more interesting question is whether there is an analog of the purely gravitational Chern-Simons invariant of \cite{Bergshoeff:2001hc}. 
This would appear to require the construction of a second 5D, $N=1/2$ superspace that would be the analog of the dilaton-Weyl formulation of Poincar\'e supergravity.

One might expect that the two different formulations could lead to different 5D, $N=1/2$ formulations based upon either old minimal or new minimal supergravity. However, this is not actually correct. As we have showed, the non-linear 5D, $N=1/2$ formulation we have introduced could be interpreted as \emph{either} old minimal or new minimal, depending on whether we choose $G$ or $H_y$ to possess non-vanishing vacuum values. This is because the underlying full supergravity involves an isotriplet $\ell^{ij}$, whose vacuum value is unfixed by equations of motion.
This suggests a more complicated mechanism is needed to describe the dilaton-Weyl formulation, which in the linearized limit resembles neither old minimal nor new minimal supergravity.

%%%%%%%%%%%%%%%%%%%%%%%%%%%%%%%%%%%%%%%%%%%%%%%%%%%%%%%%%%%%%%%%
%%%%%%%%%%%%%%%%%%%%%%%%%%%%%%%%%%%%%%%%%%%%%%%%%%%%%%%%%%%%%%%%
\section{Conclusions and Prospects}
\label{S:End}

We have given a description of pure five-dimensional, $N=1$ supergravity in terms of four-dimensional, $N=1$ superfields. 
More precisely, we are describing a 5D, $N=1/2$ superspace in which half of the supersymmetry is manifest and, in particular, off-shell. 
The remaining half of the five-dimensional supersymmetry is realized linearly but not manifestly. 
The gravitino associated to this second half sits in its own unconstrained superfield $\Psi^\alpha_i$ transforming as \eqref{E:GrinoXf}. 
To lowest order in this gravitino superfield (and for $y$-independent conformal supergravity backgrounds---cf.\ \S{}\ref{E:GrinoPunt}), the complete, non-linear action is the sum of Chern-Simons term \eqref{E:CSLagrangian} and manifestly gauge-invariant superspace volume term \eqref{E:vol}.
Besides the expected 4D, $N=1$ superspace volume density, the latter has contributions from a K\"ahler function $\sqrt{g(F)}$ with $F_y$ the field strength \eqref{E:NATHFS1}, and a non-linear function $\mathcal F$ of the tensor multiplets. 
This tensor potential function was fixed exactly by the extended, non-manifest supersymmetries. 
We checked this action by linearizing around flat space and recovering the known result \cite{Linch:2002wg}.
Finally, we used the linearized supergeometry \cite{Gates:2003qi} to construct part of the gravitational Chern-Simons action $\sim \int A\wedge R\wedge R$ in section \ref{S:R2} to cubic order in the fields of minimal 5D supergravity.

Let us highlight some noteworthy features of this construction:
\begin{enumerate}
\item There is a local conformal symmetry because of the splitting of the superspace into $\bm X$ and $Y$ parts. In particular, the physical superfields are superconformal primary fields of the 4D, $N=1$ conformal algebra. 
\item The 4D, $N=1$ supergravity theory our construction extends can be viewed alternatively as old minimal or new minimal supergravity depending on which of two compensator fields takes a vacuum value. Viewed as old minimal supergravity, it is actually a modified variant (a.k.a. 3-form supergravity) where the 3-form multiplet plays the role of the conformal compensator \cite{Linch:2002wg,Gates:2003qi}.\footnote{The linearization of this action has a hidden $Sp(4; \mathbf R)$ U-duality symmetry \cite{Linch:2015lwa}. It would be interesting to know whether this can be extended to the five-dimensional theory.}
\item It requires the full tensor hierarchy of differential $p$-forms with $p=0,\dots , 3$ even though only $p=0,1$ are physical. 
\item In the linearized old minimal form, the gravitino superfield can eat all its compensators in a Higgs-like mechanism collapsing the non-OMSG part to a sum of two squares and a radion coupling \cite{Linch:2002wg}. 
\end{enumerate}

There are many directions in which this work can be extended. 
Of course we could now couple to matter multiplets in this superspace to study phenomena in which the gravitational effects of such couplings are important ({\it e.g.}\ M-theory on Calabi-Yau 3-folds and F-theory on elliptically-fibered $G_2$ manifolds \cite{Vafa:1996xn}).
These matter fields could be either 5D or localized on lower-dimensional defects, as in \cite{Buchbinder:2003qu} (or the 5D lift of the membrane \cite{Ovrut:1997ur}). 
In the former case, one would like to understand the structure of the hyper-K\"ahler potential and the gauge kinetic function in this superspace.

The discussion of the gravitational Chern-Simons form in section \ref{S:R2} can be extended to $D>5$ dimensions. (In fact, much of the analysis of this paper was motivated by the desire to construct the supersymmetrization of the $R^4$ terms in eleven-dimensional supergravity.) This amounts to extending the derivative and all $Y$ tensors $\partial_y\to \partial_i$, $\mathbb H_y \to \mathbb H_i$, etc. for $i=1,\cdots, D-4$ and continuing the construction of the closed 4-form $(\mathbb G, \mathbb H_y)\to (\mathbb G, \mathbb H_i, \mathbb W_{\alpha ij}, \mathbb F_{ijk}, \cdots)$. 
This requires a higher-dimensional analog of reference \cite{Gates:2003qi}. 
(The requisite analysis has been carried out and will be reported elsewhere.)

The second gravitino could be incorporated into the supergeometry. 
In such a description, it cannot appear explicitly so neither can the gravitino current \eqref{E:S1}. 
The only additional explicit terms in the action would then be those involving the $X^a_y$ invariant that linearizes to \eqref{E:Xdef}. 
Such a construction can be carried out for any supergravity theory with a 4D, $N=1$ graviton and gravitino superfield, so we have chosen to present that formalism in a separate publication.

A related line of inquiry concerns a better understanding of the tensor function $\mathcal F$.\footnote{It is somewhat remarkable that the ``only'' difference between this and eleven dimensions is that there we find the reciprocal relation $\hat {\mathcal F}_\mathrm{5D} \leftrightarrow 1/\hat {\mathcal F}_\mathrm{11D}$. 
%This was foreshadowed by \eqref{E:calG} (cf.\ eq.\  4.23 of \cite{Becker:2018phr}). 
Explicitly, the eleven-dimensional version of \eqref{E:Fhat} is $\hat {\mathcal F}_\mathrm{11D} = (\bar \eta^2 \eta^2)^{1/3}$.}
It is the integral of a function that is the single-valued branch of a cubic equation \eqref{E:FhatAlg}. 
The origin of this cubic equation was elucidated by considering how 5D, $N=1$ superframe reduces to $N=1/2$ \eqref{E:ReducedFrames}, but the analysis is also reminiscent of a non-linear supersymmetry realization along the lines of \cite{Bagger:1997pi}.
Perhaps the equation \eqref{E:CompensatorAlg} can be interpreted as the lowest component of a non-linear constraint on a compensator superfield.
Such an interpretation may have far-reaching consequences for the construction of effective actions similarly to those leading to Born-Infeld theory in superspace \cite{Bagger:1994vj,Bagger:1996wp,Bagger:1997pi,Rocek:1997hi}.

These last three points are related by the observation that the gravitational Chern-Simons action is not invariant under the extended linearized supergravity transformations because the graviphoton field shifts under \eqref{E:NATHgauge1}.
This is expected since higher-derivative corrections to the supersymmetry transformations are known to be required, even 
in completely off-shell formulations (see \cite{Bergshoeff:1986jm} for an example already in higher-derivative Yang-Mills theory).
In the aforementioned covariant supergeometrical formulation, the 2-derivative action must transform under the corrections so as to cancel the shift of the gravitational Chern-Simons term. 
Assuming this can be done, the Chern-Simons term is still not expected to be invariant under the higher-derivative correction to the supersymmetry transformations.
This line of reasoning leads to ever-higher corrections to the supersymmetry transformations and the action, just as it does in the familiar component analysis. 
However as mentioned above, the {\em form} of the action is fixed by the off-shell part of the supersymmetry to be an $X^a_y$-corrected version of that found in \S{}\ref{S:Action}.
We do not currently have an adequate understanding of how this tension is resolved, but (by analogy to the Witten anomaly in 11D \cite{Witten:1996md}) we expect that part of the solution involves a shift of the tensor hierarchy field strengths by terms quadratic in the curvature tensor.
A primary motivation for constructing this 5D, $N=1/2$ formalism is to use it to study this question in a simplified (relative to 11D) framework.

% Additional stuff

% 
% It would be interesting to carry out a more careful calculation to determine the precise form of $\hat U$, its non-manifest supersymmetry transformation, and the precise form of the equation preventing the closure. 
% We reiterate, however, that the analogous analysis is not expected to be possible in formulations in which there is no off-shell superspace ({\it e.g.}\ 11D supergravity). 
% 

%%%%%%%%%%%%%%%%%%%%%%%%%%%%%%%%%%%%%%%%%%%%%%%%%%%%%%%%%%%%%%%%%
%%%%%%%%%%%%%%%%%%%%%%%%%%%%%%%%%%%%%%%%%%%%%%%%%%%%%%%%%%%%%%%%%
\section*{Acknowledgements}
It is a pleasure to thank our referees for an exceptionally thorough report that inspired us to extend the results and improve the presentation in version 2 of this paper. 
This work is partially supported by National Science Foundation grants PHY-1820921 and PHY-1820912 and the Mitchell Institute for Fundamental Physics and Astronomy at Texas A\&M University.

\appendix
%%%%%%%%%%%%%%%%%%%%%%%%%%%%%%%%%%%%%%%%%%%%%%%%%%%%%%%%%%%%%%%%
%%%%%%%%%%%%%%%%%%%%%%%%%%%%%%%%%%%%%%%%%%%%%%%%%%%%%%%%%%%%%%%%
\section{Derivations}
\label{S:Derivations}
In this appendix, we provide some details of the derivation of the results in section \ref{S:Action}. 
These derivations follow those of reference \cite{Becker:2018phr} rather closely, so we will be schematic in places where more rigor obscures the presentation. 
Specifically, we specialize to trivial 4D, $N=1$ conformal supergravity backgrounds, and, throughout this calculation, we suppress the $y$ indices as well as spinor indices. Juxtaposed spinors are contracted with the suppressed chiral indices up-to-down ($\mathcal D \Psi = \mathcal D^\alpha \Psi_\alpha$) and down-to-up for antichiral ($\bar {\mathcal D} \bar \Psi = \bar {\mathcal D}_{\dt \alpha} \bar \Psi^{\dt \alpha}$). 

The non-linear gravitino transformation 
\eqref{E:GrinoXf} can be rewritten as\footnote{The eleven-dimensional theory had a quadratic term $W_\alpha \bar \Omega$ but that is ruled out here by 5-charge.}
\begin{align}
%\label{E:GrinoXf}
\delta_0 \Psi_\alpha 
&= \Xi_\alpha + (\bar GG)^{-1/3} \mathcal G {\mathcal D}_\alpha ( \hat \Omega + i \check \Omega)
~.
\end{align}
There is a gauge-for-gauge symmetry
\begin{align}
\label{E:g4g}
\delta \Omega = \bar \phi 
~~~\textrm{with}~~~
\bar {\mathcal D}_{\dt \alpha} \phi =0 
\end{align}
which played an important role in the eleven-dimensional theory. 
In this case it is less powerful but it suffices to rule out a transformation of the form
$\delta \Phi \sim \bar {\mathcal D}^2 (\bar \Omega \mathcal U)$ for some $\mathcal U$ \cite{Becker:2018phr}.
The transformation of the gravitino-current coupling that we will cancel is
\begin{align}
\label{E:deltaS1}
\delta_0 S_1&= \frac1{\kappa^2} \int d^5x \int d^4 \theta  \, \left[ 
	\Xi J
	+i {\mathcal D}  \check \Omega  (\bar GG)^{-1/3} \mathcal G J
	\right]
	+ \mathrm{h.c.}
\end{align}
(In particular, we will ignore the $\hat \Omega$ part; we can imagine that we are gauging away the KK field (cf.\ \S\ref{S:HiggsedAction}).)

To compute the transformations of the volume functional \eqref{E:vol}, it is convenient to first rewrite it as
\begin{align}
\label{E:vol2}
S_{vol} &= - \frac3{\kappa^2} \int d^5x \int d^4\theta   \, H {\mathcal H}(h)
~~~\textrm{with}~~~
{\mathcal H}(h) := h^{-1}{\mathcal F}(h)
\cr
\Rightarrow~~~
\delta S_{vol} &= 
	- \frac3{\kappa^2} \int d^5x \int d^4\theta   \, \left[ \delta H {\mathcal H} + H h {\mathcal H}' \delta \log h
	\right]
~.
\end{align}
For the Chern-Simons action, the variation is 
\begin{align}
\label{E:deltaCSA}
\kappa^2 \delta L_{CS} = \frac {3i}4 \int d^2 \theta  \, \delta \Phi W^\alpha W_\alpha + \textrm{h.c.} 
	+ 3 \int d^4 \theta   \, \delta V \omega(W, F)
~.
\end{align}
(That the ``abelian'' part just gives a factor of 3 follows from the fact that this action is the superspace analog of $\int A dAdA$; we ignore the non-abelian correction.)

Next we compute the actual transformations of these terms. 
The field strengths \eqref{E:NATHFS} satisfy $\delta_0(\textrm{anything})=0$ by definition, and transform under the extended supersymmetry parameters
\begin{subequations}
\label{E:d1NATH}
\begin{align}
\delta_1 F &= \frac1{2i} (\Xi W - \bar \Xi \bar W) - {\partial_y} (\check \Omega F)
\\
\delta_1 W &= -\frac14 \bar {\mathcal D}^2 {\mathcal D} (\check \Omega F) - (\Xi \mathcal W) W
\\
\delta_1 H &= -\frac1{2i}\left[ {\mathcal D} (G \Xi) - \bar {\mathcal D} (\bar G \bar \Xi)\right] - {\partial_y} (\check \Omega H)
\\
\delta_1 G &= -\frac14 \bar {\mathcal D}^2 (\check \Omega H) - G\Xi \mathcal W
\\
\delta_1 \mathcal W &= -\frac14 \bar {\mathcal D}^2 {\mathcal D} \hat \Omega
~.
\end{align}
\end{subequations}
In the following two subsections, we will split up the calculation into the $\Xi$ part and the $\Omega$ part.

%%%%%%%%%%%%%%%%%%%%%%%%%%%%%%%%%%%%%%%%%%%%%%%%%%%%%%%%%%%%%%%%
\subsection{Invariance under $\Xi$: Supercurrent}
\label{S:XiInv}

We compute for $\Xi$
\begin{align}
\label{E:htransX}
\delta_{\Xi} \log h  
	&= \frac i2 F^{-1} \Xi W  - \frac 13  \Xi {\mathcal W} + \frac i2 H^{-1}  {\mathcal D} (G \Xi )
\end{align}
so that under $\Xi$ the volume functional changes by
\begin{align}
\delta_{\Xi} S_{vol} &= 
	- \frac3{\kappa^2} \int d^5x \int d^4\theta   \, \left[ \delta_{\Xi} H {\mathcal H} + H h {\mathcal H}' \delta_{\Xi} \log h
	\right]
\\
	& = {\kappa^2} \int d^5x \int d^4\theta   \, \Xi
	\left[
	-\frac{3i}2 W F^{-1} H h \mathcal H'
	+ \mathcal W H h \mathcal H'
	+\frac{3i}2 {\mathcal D} ( \mathcal H + h \mathcal H' )
	\right]
\cr
	& = {\kappa^2} \int d^5x \int d^4\theta   \, \Xi
	\left[
	(\bar GG)^{1/3} \hat{\mathcal F} \left(\frac{3i}2  W -  F \mathcal W \right)
	+\frac{3i}2 G {\mathcal D} {\mathcal F}'
	\right]
~,	
\nonumber	
\end{align}
where we used that $h^2 \mathcal H' = -\hat {\mathcal F}$.
The Chern-Simons action is $\Xi$-invariant since it is independent of $\Sigma$ and the $\Phi W^2$ term transforms into $W^3 \equiv 0$ (cf.\ \ref{E:deltaCSA}). 
Therefore, the variation above must be canceled by the variation \eqref{E:deltaS1} of the gravitino-current coupling.
This fixes the current to
\begin{align}
\label{E:J2}
J &=\frac{3i}2 W F^{-1} H h \mathcal H'
	- \mathcal W H h \mathcal H'
	-\frac{3i}2 {\mathcal D} ( \mathcal H + h \mathcal H' )
\cr
&= - (\bar GG)^{1/3} \left[\frac {3i}2 W- \mathcal WF \right] \hat {\mathcal F} 
	- \frac {3i}2 G{\mathcal D} {\mathcal F}'
\cr	
&= \left[  \frac {i}2 W {\partial \over \partial F} 
	- \mathcal W {\partial  \over \partial \log(\bar GG)} 
	+ \frac {i}2 G{\mathcal D} {\partial \over \partial H} 
	\right] \left(-3 (\bar G G)^{1/3} F {\mathcal F}\right)
~	
\end{align}
up to terms that are $\bar {\mathcal D}$ of some 4-vector with $(\Delta, w, d, q) = (3, 0, 0, 0)$, as these are in the kernel of the $\Xi$ transformation. 
Studying currents made only from the tensor hierarchy field strengths (for gauge invariance) and their covariant derivatives, one concludes that there are no such correction terms.

%%%%%%%%%%%%%%%%%%%%%%%%%%%%%%%%%%%%%%%%%%%%%%%%%%%%%%%%%%%%%%%%
\subsection{Invariance under $\Omega$}
\label{S:OmegaInv}
Now that we have the gravitino current, we can determine the unknown function $\mathcal G$ in the $\Omega$ part of the gravitino transformation \eqref{E:GrinoXf}.
We use the observation that under \eqref{E:d1NATH}
\begin{align}
\label{E:htransO}
\delta_{\check \Omega} h  
	&= - \check \Omega {\partial_y} h + 
	\left[
	\frac1{12 G} \bar {\mathcal D}^2 \check \Omega H
	+ \frac1{6 G} \bar {\mathcal D} \check \Omega \bar {\mathcal D} H
	+\mathrm{h.c.}
	\right]
~,	
\end{align}
where we have used \eqref{E:NATHBIH}. 
From this, we obtain
\begin{align}
\delta_{\check \Omega} S_{vol} 
	&= 
	- \frac3{\kappa^2} \int d^5x \int d^4\theta   \, \left[ \delta_{\check \Omega} H {\mathcal H} + H {\mathcal H}' \delta_{\check \Omega} h
	\right]
\cr	
&= 
	- \frac3{\kappa^2} \int d^5x \int d^4\theta   \, \Bigg[ \delta_{\check \Omega} H {\mathcal H} 
	- \check \Omega H {\partial_y} h {\mathcal H}' 
\cr&	\hspace{35mm} + H {\mathcal H}' \left(
	\frac1{12 G} \bar {\mathcal D}^2 \check \Omega H
	+ \frac1{6 G} \bar {\mathcal D} \check \Omega \bar {\mathcal D} H
	+\mathrm{h.c.}
	\right)
	\Bigg]
%\cr	
\nonumber
\end{align}
\begin{align}
	&=- \frac3{\kappa^2} \int d^5x \int d^4\theta   \, 
	\frac1{12 G} \bar {\mathcal D} (\bar {\mathcal D} \check \Omega H^2) h {\mathcal H}'
+\mathrm{h.c.}	
\cr
	&= \frac 1{4\kappa^2} \int d^5x \int d^4\theta   \, \bar G^{-1} H^2 {\mathcal D} \check \Omega {\mathcal D}(h {\mathcal H}')
+\mathrm{h.c.}	
\end{align}
In particular, the $\check \Omega$ part ({\it i.e.}\ without derivatives) cancels, consistent with the observation that such a term would not be invariant under the gauge-for-gauge symmetry \eqref{E:g4g}.
The Chern-Simons term contributes
\begin{align}
\kappa^2 \delta_\Omega L_{CS} = 
	3 \int d^4 \theta   \, \check \Omega F \left[ W{\mathcal D} F +\tfrac12 F {\mathcal D} W 
	+\mathrm{h.c.}
		\right]
\cr
= 
	\frac32 \int d^4 \theta   \, \check \Omega {\mathcal D} (F^2 W)
	+\mathrm{h.c.}
\end{align}
for a total of 
\begin{align}
\delta_{\check \Omega} S_0 
	&= \frac 1{\kappa^2} \int d^5x \int d^4\theta   \, {\mathcal D} \check \Omega
	\left[
	-\frac32 F^2 W
	+\frac14 \bar G^{-1} H^2  {\mathcal D}(h {\mathcal H}')
	+O(\mathcal W)
	\right]
+\mathrm{h.c.}	
\end{align}
Comparing to the gravitino transformation \eqref{E:deltaS1} with the current given by \eqref{E:J2}, we find 
\begin{align}
-\frac32 F^2 W
	+\frac14 \bar G^{-1} H^2  {\mathcal D}(h {\mathcal H}')
	+O(\mathcal W)
&=- i (\bar GG)^{-1/3} \mathcal G J 
\\
&= \left[-\frac {3}2 W - i\mathcal WF \right] {\mathcal G }\hat {\mathcal F} 
	- \frac {3}2 G(\bar GG)^{-1/3} {\mathcal G} {\mathcal D} {\mathcal F}'
~,	
\nonumber
\end{align}
whence we read off $\mathcal G = \hat {\mathcal F}^{-1} F^2$ and $h^2  {\mathcal D}(h {\mathcal H}') = -6 \hat {\mathcal F}^{-1} {\mathcal D} {\mathcal F}'$.\footnote{This agrees with the result from 11D in which $\mathcal G$ is a Hermitian bi-linear form. Here the imaginary anti-symmetric part is absent since there cannot be such a form in co-dimension 1.
This also has the correct limit $\vev{\mathcal G} = 1$. 
} %end footnote
The coefficient of ${\mathcal D} h$ of this equation implies $h^3 (h{\mathcal H}')' = 6  \log'\hat{\mathcal F}$.
Note that the left-hand side scales like $h$ while the right-hand side scales like $h^{-1}$.
Using 
$\mathcal F = h {\mathcal H} $, 
$\mathcal F' = {\mathcal H} + h {\mathcal H}'$, 
$\hat {\mathcal F} 
={\mathcal F} - h {\mathcal F}'
= - h^2{\mathcal H}'$, 
and ${\mathcal F}'' = - h^{-1} \hat{\mathcal F}'$, 
we find that 
\begin{align}
h \hat {\mathcal F}( \hat {\mathcal F} - h \hat {\mathcal F}') = \frac\alpha2  \hat{\mathcal F}'
~~~\textrm{with}~~~
\alpha = 12
~.
\end{align}
Since $\mathcal F$ is really a function of $h^2$, this is equivalent to \eqref{E:calFdiffEQ} considered as a function of $x$.
The latter can be integrated to the form\footnote{In eleven dimensions, the analogous equation is $\left( 1 + 3x \hat {\mathcal F}\right)\hat{\mathcal F}^{3} - 1 = 0$. (Our $x$ differs from that of \cite{Becker:2018phr} by a factor of 12.) 
} %end footnote
\begin{align}
%\label{E:FhatAlg}
\left( 1 + 3 x \hat {\mathcal F}\right)\hat{\mathcal F}^{-3} - 1 = 0
~,
\end{align}
as is easily checked by differentiating and clearing negative powers. 
Thus, we obtain \eqref{E:FhatAlg}.
% 
% The constant of integration has been fixed by matching to $x\to 0$ (${\mathcal F}(0) = 1$ $\Rightarrow$ $\hat {\mathcal F}(0) =1$).
% 
The principal branch is given by
\begin{align}
\hat{\mathcal F} 
&= 
{ 2^{2/3} x  +  \left(1 + \sqrt{ 1 - 4 x^3}
	\right)^{2/3}
\over
2^{1/3}\left(
	1 + \sqrt{1 - 4x^3 }
	\right)^{1/3}}
\cr
&= 1 + x -\frac {x^3}3 +\frac {x^4}3 -\frac {4x^6}9 +\frac {5x^7}9 
-\frac {77x^9}{81} +\frac {104 x^{10}}{81} +O(x^{12})
\end{align}
\begin{figure}[t]
\center
\scalebox{0.40}{
\includegraphics{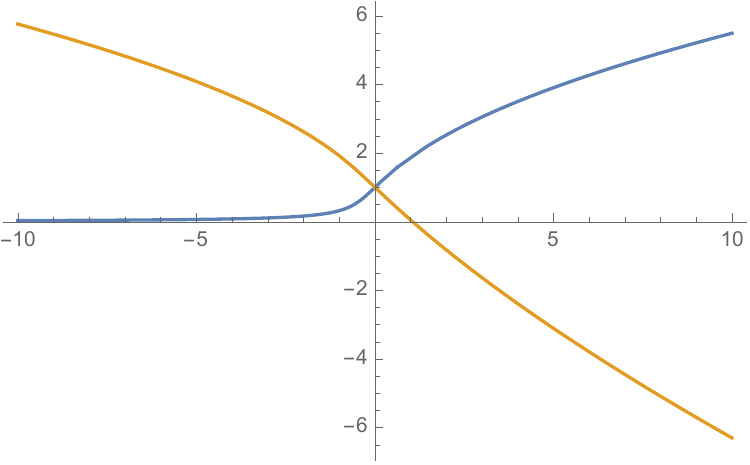}
}
\begin{caption}{
The single-valued branch of the function $\hat {\mathcal F}(x) = {\mathcal F} - 2 x {\mathcal F}'$ and ${\mathcal F}$ (choosing the integration constant $c_1=0$).
The former is the positive definite one.
The latter vanishes at $x\approx 1.06$ before running off to $-\infty$ as $-\sqrt{3x}\log \sqrt{x}$ (cf.\ \ref{E:FpertLarge}).
}
\label{F:FhatF}
\end{caption}
\end{figure}To find $\mathcal F$ from this, we must solve the defining equation \eqref{E:FhatDef}.
Had $\hat{\mathcal F}$ been convex, this would just be the Legendre transform. Instead, we find the complicated solution plotted in figure \ref{F:FhatF}.

We emphasize that the function $\hat {\mathcal F}$ and its integrated form $\mathcal F$ must be real for $x \geq 0$, as all non-negative values of $x$ are physically permissible. The form given above is only manifestly real for $4x^3 \leq 1$. $\hat {\mathcal F}$ may equivalently be written 
\begin{align}
\hat{\mathcal F} = \frac{1}{2^{1/3}} \Big(
    (1 - i \sqrt{4x^3-1})^{1/3} + (1+i \sqrt{4x^3-1})^{1/3}
\Big)
\end{align}
which is manifestly real for $4x^3\geq 1$.

%%%%%%%%%%%%%%%%%%%%%%%%%%%%%%%%%%%%%%%%%%%%%%%%%%%%%%%%%%%%%%%%
%%%%%%%%%%%%%%%%%%%%%%%%%%%%%%%%%%%%%%%%%%%%%%%%%%%%%%%%%%%%%%%%
{\footnotesize
%\bibliography{/Users/wdlinch3/Dropbox/Rashoumon/LaTeX/BibTex/BibTex}
\bibliography{./Simple5Draftv2.bbl}

\bibliographystyle{unsrt}

}

\end{document}